\documentclass[aps,amsmath,amssymb,reprint,superscriptaddress]{revtex4-2}

\usepackage{graphicx}
\usepackage{dcolumn}
\usepackage{bm}
\usepackage{ulem} 
\usepackage{siunitx} 
\usepackage{hyperref}
\usepackage{xcolor}

\usepackage{float}
\usepackage{placeins}
\newcommand{\hl}[1]{{\color{black} #1}}

\begin{document}

\preprint{APS/123-QED}

\title{Magnomechanical backaction corrections due to coupling to higher-order Walker modes and Kerr nonlinearities}

\author{V.A.S.V. Bittencourt}%
\email{sant@unistra.fr}
\affiliation{ISIS (UMR 7006), Universit\'{e} de Strasbourg, 67000 Strasbourg, France}
\affiliation{Max Planck Institute for the Science of Light, Staudtstr. 2, PLZ 91058 Erlangen, Germany}

\author{C.A. Potts}
\affiliation{Kavli Institute of NanoScience, Delft University of Technology, PO Box 5046, 2600 GA Delft, Netherlands}
\affiliation{Department of Physics, University of Alberta, Edmonton, Alberta T6G 2E9, Canada}

\author{Y. Huang}
\affiliation{Department of Physics, University of Alberta, Edmonton, Alberta T6G 2E9, Canada}

\author{J.P. Davis}
\affiliation{Department of Physics, University of Alberta, Edmonton, Alberta T6G 2E9, Canada}

\author{S. {Viola Kusminskiy}}
\email{kusminskiy@physik.rwth-aachen.de}
\affiliation{Institute for Theoretical Solid State Physics, RWTH Aachen University, 52074 Aachen, Germany}
\affiliation{Max Planck Institute for the Science of Light, Staudtstr. 2, PLZ 91058 Erlangen, Germany}

\date{\today}

\begin{abstract}
The radiation pressure-like coupling between magnons and phonons in magnets can modify the phonon frequency (magnomechanical spring effect) and decay rate (magnomechanical decay) via dynamical backaction. Such effects have been recently observed by coupling the uniform magnon mode of a magnetic sphere (the Kittel mode) to a microwave cavity. In particular, the ability to evade backaction effects was demonstrated [C.A. Potts \textit{et al.}, arXiv:2211.13766 [quant-ph] (2022)], a requisite for applications such as magnomechanical-based thermometry. However, deviations were observed from the predicted magnomechanical decay rate within the standard theoretical model. In this work, we account for these deviations by considering corrections due to (i) magnetic Kerr nonlinearities and (ii) the coupling of phonons to additional magnon modes. Provided that such additional modes couple weakly to the driven cavity, our model yields a correction proportional to the average Kittel magnon mode occupation. We focus our results on magnetic spheres, where we show that the magnetostatic Walker modes couple to the relevant mechanical modes as efficiently as the Kittel mode. Our model yields excellent agreement with the experimental data.
\end{abstract}

\maketitle

The dipolar interaction between the magnetization and microwaves confined in a cavity can yield strong coupling between magnons (quanta of spin waves) and microwave photons. After the first theoretical predictions of the strong magnon-microwave coupling \cite{soykal2010strongfield}, cavity magnonic systems consisting of a magnetic element loaded in a microwave cavity were realized in different architectures \cite{huebl2013high, zhang2014strongly, tabuch2014hybridizing, goryachev2014highcooperativity, lambert2016cavitymediated, morris2017strong, potts2020strong}. The unique tunability of magnons combined with the ability to drive and read out the microwave cavity makes such systems a promising platform for several applications \cite{lachance2019hybrid, li2020hybridmagnonics, awschalom2021quantumengineering, chumak2021roadmap,rameshti2022cavity}, such as the generation of squeezed and entangled states \cite{elyasi2020resources, nair2020deterministicquantum}, the indirect coupling to qubits to detect and manipulate magnons \cite{tabuchi2015coherent,lachance2017resolvingquanta,lachance2020entanglementbased,sharma2022protocolforgenerating,kounalakis2022analog}, and sensing of magnetic fields \cite{flower2019broadening, crescini2020axionsearch, ebrahimi2021singlequadrature, ikeda2022axionsearch}.

Magnons can also couple to other degrees of freedom, opening the opportunity of probing and manipulating these via their coupling to the hybridized magnon-microwave polaritons \cite{rameshti2022cavity}. In particular, magnetoelastic effects \cite{kittel1949physicaltheory, landau1984electrodynamics,callen1968magnetostriction, gurevich2020magnetization} couple the magnetization and the mechanical vibrations of a magnetic material, yielding an interaction between magnons and phonons \cite{zhang2016cavity}. Such magnomechanical coupling can be either resonant or parametric \cite{Gonzalez_2020_Theory}. The resonant coupling is relevant for specific geometries where certain magnon modes are resonant with the elastic vibrations of the medium, for example, for magnetic spheres with radii ranging from $\sim 10$ nm to $\sim 10$ $\mu$m \cite{Gonzalez_2020_Theory} and in magnetic films \cite{an2020coherentlongrange, litvinenko2021tunablemagnetoacoustic,schlitz2022magnetizationdynamics}. The second type of coupling, parametric coupling, is relevant for geometries in which the magnon frequency is far detuned from the phonon, such as for micrometer-sized magnetic spheres \cite{zhang2016cavity,potts2021dynamical}. The interaction Hamiltonian resembles the radiation pressure coupling between phonons and photons commonly found in optomechanical systems \cite{aspelmeyer2014cavity}. When magnons are driven, the magnomechanical interaction is enhanced and the driven-dissipative dynamics of the coupled system result in dynamical backaction on the vibrational modes. Specifically, the phonon frequency and decay rate are modified, referred to as the magnomechanical spring effect and magnomechanical decay, respectively \cite{zhang2016cavity, potts2020magnon, potts2021dynamical}.

Dynamical backaction is the basis of several proposed applications of magnomechanical systems, from state preparation and generation of entangled states \cite{li2018magnon, nair2020deterministicquantum, cheng2021tripartite, sarma2021cavity, li2021entangling}, to effects that are closely related to the optomechanical counterpart, such as magnomechanical sideband cooling and amplification of phonons \cite{ding2020goundstatecooling, ding2019phononlaser}, albeit operating in the microwave regime. Moreover, cavity magnomechanical systems provide a unique tunability of dynamical backaction due to the hybridization between magnons and microwaves, which can be used to fulfill a triple resonance condition by tuning an external bias magnetic field. Dynamical backaction was first probed in magnomechanics in a system consisting of a magnetic sphere of yttrium iron garnet (YIG) loaded into a three-dimensional microwave cavity \cite{zhang2016cavity}. More recent experiments have demonstrated the full array of dynamical backaction effects in these systems \cite{potts2021dynamical} and demonstrated the capability of avoiding the induced magnomechanical decay \cite{potts2022dynamicalbackaction}. Dynamical backaction evasion can enable the application of cavity magnomechanics in thermometry \cite{potts2020magnon}, where, similar to proposals and experiments in optomechanical systems \cite{borkje2010observability,purdy2017quantum}, the phonon mode should be neither cooled nor heated by the drive.

Experiments and proposals for cavity magnomechanical systems have so far focused on the coupling to a single magnon mode, the uniform precession of the magnetization called the Kittel mode. Nevertheless, a magnetic sphere supports a whole set of magnon modes called Walker modes \cite{walker1958resonant,fletcher1959ferrimagnetic} which can also couple to a given vibration mode, in principle even stronger than the Kittel mode. Weak inhomogeneities in the microwave field can drive such higher-order magnon modes, modifying the backaction effects. \hl{Furthermore, magnon nonlinearities due to crystalline anisotropy \cite{maconald1951ferromagnetic,stancil2009spinwaves} can also affect dynamical backaction. Nonlinearities appear as self- and cross-Kerr terms in the magnomechanical Hamiltonian, inducing static frequency shift of the modes which have been reported in Refs.\,\cite{wang2016magnonkerr, shen2022mechanical}. Additionally, the magnon Kerr nonlinearity can yield a bistable behavior of the magnons \cite{zhang2019theoryofthe, pan2022bistability} and mechanics \cite{shen2022mechanical}, and induce quantum phase transitions \cite{zhang2021ptbreaking}, among other phenomena. Nevertheless, no description of how such nonlinearities affect dynamical backaction beyond the static frequency shift has been reported. For instance, in optomechanical systems, effects of the cavity Kerr nonlinearity in an electromechanical system  under sideband cooling have been recently observed \cite{zoepfl2022kerrenhanced}.}

\begin{figure}[h]
\includegraphics[width = 0.5\textwidth]{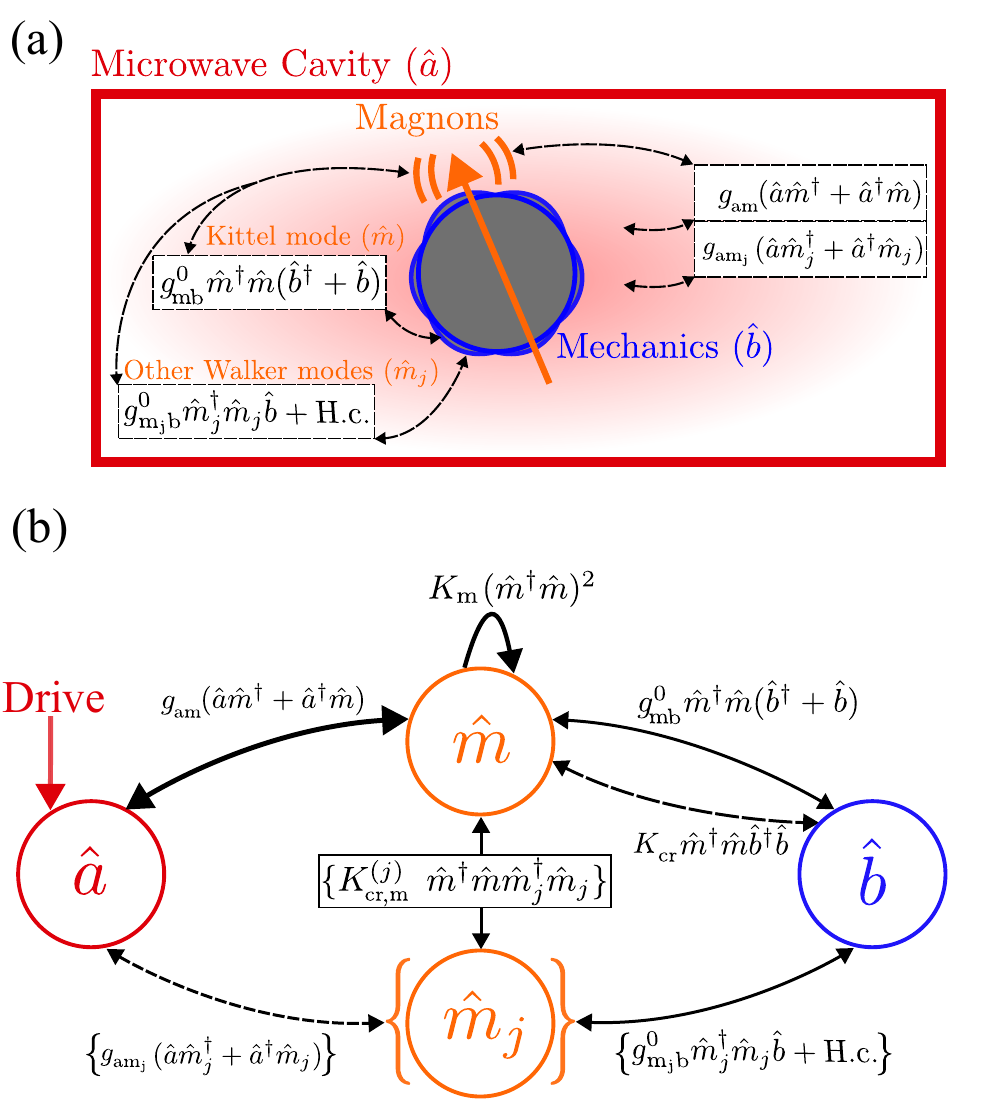}
\caption{(a) Schematic representation of the cavity magnomechanical system. A magnetic element is loaded in a microwave cavity. The magnon modes couple resonantly with a mode from the cavity and parametrically with the mechanical vibrations of the magnet. (b) Schematics of the model describing the cavity magnomechanical system with several magnon modes coupled to the same phonon mode [see Hamiltonian \eqref{App02:Eq01}]. Our model assumes that only one of the magnon modes, denoted by $\hat{m}$, couples strongly to the cavity mode. The red arrow indicates the microwave drive.}
\label{Fig:App02}
\end{figure}

In this work, we extend the theory of dynamical backaction in cavity magnomechanical systems to include Kerr nonlinearities and the coupling of a phonon mode to several magnon modes. We consider the framework depicted in Fig. \ref{Fig:App02}, where a microwave cavity mode couples strongly to a magnon mode and weakly to a set of additional magnon modes. Those in turn exhibit nonlinearities and interact via a radiation pressurelike coupling to a single phonon mode. We derive the phonon self-energy, describing the frequency shift and the magnomechanical decay rate, generalizing previous results \cite{potts2020magnon,potts2021dynamical}. The overall effect of the coupling to the additional magnon modes is a correction proportional to the average number of Kittel magnons. We evaluate our model for the case of a magnetic sphere, computing numerically the coupling rates between the (magnetic) Walker modes and the mechanical mode probed in Ref.\,\cite{potts2022dynamicalbackaction}. At low driving powers our model introduces corrections that agree well with the measured data in Ref.\,\cite{potts2022dynamicalbackaction}, explaining the observed shift in the magnomechanical decay rate. At higher driving powers, there are further deviations which are not captured by our model. These are, however, only relevant for driving frequencies detuned from the backaction evasion point.

This paper is organized as follows. In Sec.\,\ref{Sec:PhononSelfEnergy} we present a brief review of the description of dynamical backaction in cavity magnomechanics for the cases described in the literature, e.g., Refs.~\cite{zhang2016cavity,potts2021dynamical}. In Sec.\,\ref{Sec:AdditionalModesSelfEnergy} we include Kerr nonlinearities and the coupling to weakly driven additional magnon modes, and the phonon self-energy. Since those corrections depend on how strongly the additional magnon modes couple to the phonon mode, we specialize further our model to a magnetic sphere geometry as probed in Ref.\,\cite{potts2022dynamicalbackaction}. In Sec.  \ref{Sec:MagnoMechCoupling}, we briefly review the derivation of the magnomechanical coupling following the literature \cite{Gonzalez_2020_Theory}, and in subsection \ref{Sec:MagnoMechCouplingSphere} we use the model to numerically evaluate the coupling between Walker modes with frequencies in a small range around the Kittel mode frequency and a relevant mechanical mode of a magnetic sphere. In Sec.\,\ref{Sec:Evaluation}, we compare our model to the experimental results presented in Ref. \cite{potts2022dynamicalbackaction} and show that our generalized theory quantitatively accounts for the observed magnomechanical decay for a large range of parameters. Finally, in Sec.\,\ref{Sec:Conclusions} we present our conclusions.

\section{Phonon self-energy and dynamical backaction evasion}
\label{Sec:PhononSelfEnergy}

The dynamics of a cavity magnomechanical system consisting of a microwave mode ($\hat{a}$ with frequency $\omega_{\rm{a}}$) and a magnon mode ($\hat{m}$ with frequency $\omega_{\rm{m}}$) coupled parametrically to a phonon mode ($\hat{b}$ with frequency $\Omega_{\rm{b}}$) is described by the Hamiltonian \cite{zhang2016cavity}
\begin{equation}
\label{Eq:1Hamiltonian}
\begin{aligned}
\frac{\hat{\mathcal{H}}}{\hbar} &= \omega_{\rm{a}} \hat{a}^\dagger \hat{a} + \omega_{\rm{m}} \hat{m}^\dagger \hat{m} + \Omega_{\rm{b}} \hat{b}^\dagger\hat{b}\\
& \quad +g_{\rm{am}} \left( \hat{a} \hat{m}^\dagger + \hat{a}^\dagger \hat{m} \right) + g_{\rm{mb}}^0 \hat{m}^\dagger \hat{m} \left( \hat{b}^\dagger + \hat{b} \right) \\
& \quad + i \sqrt{\kappa_e} \epsilon_{\rm{d}}\left(\hat{a} e^{i \omega_{\rm{d}} t} - \hat{a}^\dagger e^{-i \omega_{\rm{d}} t} \right).
\end{aligned}
\end{equation}
The magnon-microwave coupling rate $g_{\rm{am}}$ is due to a magnetic dipole interaction between the ferromagnetic resonance of the material and the microwave cavity. The parametric magnon-phonon coupling, with the single magnon coupling rate $g_{\rm{mb}}^0$, is due to magnetoelastic effects. The last term in Eq.~\eqref{Eq:1Hamiltonian} describes the coherent drive of the microwave cavity at a frequency $\omega_{\rm{d}}$ with an amplitude $\epsilon_{\rm{d}}=\sqrt{\mathcal{P}/\hbar \omega_{\rm{d}}}$, where $\mathcal{P}$ is the drive power and $\kappa_e$ is the decay rate of the cavity to the external drive port.

In the weak magnomechanical coupling limit, both the phonon frequency $\Omega_{\rm{b}}$ and the decay rate $\Gamma_{\rm{b}}$ are modified by the coupling to the driven magnons. The respective shifts are given by
\begin{equation}
\label{Eq:Modifics}
\begin{aligned}
\delta \Omega_{\rm b} &= - {\rm{Re}}[\Sigma[\Omega_{\rm{b}}]], \\
\Gamma_{\rm{mag}} &= 2 {\rm{Im}}[\Sigma[\Omega_{\rm{b}}]],
\end{aligned}
\end{equation}
where $\Sigma[\omega]$ is the phonon self-energy, obtained by analyzing the linearized dynamics of the system \cite{potts2020magnon, potts2021dynamical}, and reads
\begin{equation}
    \Sigma[\omega] = i\vert g_{\rm mb} \vert^2 ( \Xi[\omega] - \Xi^*[-\omega]).
    \label{Eqn:06}
\end{equation}
From now on we refer to $\delta \Omega_{\rm b}$ as the magnomechanical frequency shift and to $\Gamma_{\rm{mag}}$ as the magnomechanical decay rate.
Here, $g_{\rm mb} = g^0_{\rm mb}\langle \hat{m}\rangle$ is the enhanced magnomechanical coupling rate, with $\vert \langle \hat{m} \rangle\vert^2$ the steady-state magnon population. The function $\Xi[\omega]$ is a modified Kittel mode susceptibility given by
\begin{equation}
\label{Eq:XiKit}
\Xi^{-1}[\omega] = \chi_{\rm m}^{-1}[\omega] + g_{\rm am}^2\chi_{\rm a}[\omega]
\end{equation}
which depends on the magnon susceptibility $\chi_{\rm{m}}[\omega] = [-i(\Delta_{\rm{m}}+\omega) + \gamma_{\rm{m}}/2]$, and the microwave susceptibility $\chi_{\rm{a}}[\omega] = [-i(\Delta_{\rm{a}}+\omega) + \kappa/2]$. The detuning between the microwave (magnon) mode and the drive is $\Delta_{\rm{a(m)}} = \omega_{\rm{d}}-\omega_{\rm{a(m)}}$. $\gamma_{\rm{m}}$ is the magnon decay rate and $\kappa$ the total microwave decay.

The value and the sign of the magnomechanical decay rate depend on the drive frequency, which can tune scattering processes that upconvert or downconvert excitations in the system. Depending on the drive, it is possible to make one of such processes more efficient than the other, yielding a positive (cooling) or negative (amplification) magnomechanical decay rate, in a situation akin to what is found in standard optomechanical systems \cite{aspelmeyer2014cavity, wilson2007theoryofgroundstate, marquardt2007quantumtheory}. Different from optomechanics, in cavity magnomechanical systems magnons and microwaves hybridize, yielding the unique situation where the different scattering processes that contribute to dynamical backaction are associated with mechanical sidebands of hybridized modes \cite{zhang2016cavity,potts2022dynamicalbackaction}. The hybrid magnon-microwave modes have frequencies $\omega_{\pm }$ that are separated by
\begin{equation}
\label{Eq:HybridModeFrequencies}
\omega_+ - \omega_- = \sqrt{4 g_{\rm{am}}^2 + (\omega_{\rm{a}} -\omega_{\rm{m}})^2}.
\end{equation}
We will refer to the mode with frequency $\omega_+$ as the upper hybrid mode, and the mode with frequency $\omega_-$ as the lower hybrid mode.
When the microwave drive is set at a frequency between $\omega_+$ and $\omega_-$, the scattering from the blue sideband of the lower hybrid mode can be balanced by the scattering to the red sideband of the upper hybrid mode, yielding dynamical backaction evasion: the magnomechanical decay rate vanishes. Consequently, the drive at which dynamical backaction evasion happens can be obtained from the condition
\begin{equation}
\Gamma_{\rm{mag}} = 0.
\end{equation}
In a system satisfying the two-phonon triple resonance condition $\omega_+ - \omega_- = 2 \Omega_{\rm b}$, and for resonant magnons and microwaves, such drive frequency is exactly at $(\omega_+ + \omega_-)/2$ \cite{potts2022dynamicalbackaction}. The ability to tune a magnomechanical system in the dynamical backaction evasion regime was recently demonstrated in Ref.\,\cite{potts2022dynamicalbackaction}, and is a requirement for implementing a magnomechanical-based primary thermometer \cite{potts2020magnon}.
 
Equation\,\eqref{Eqn:06} only takes into account the interaction of a phonon mode with a single magnon mode. However, multiple magnetostatic modes can couple to a given phonon mode \cite{Gonzalez_2020_Theory}, modifying the magnomechanical decay rate. The different scattering processes to and from the additional magnomechanical sidebands can thus change the frequency at which dynamical backaction is evaded. For instance, in the experimental data shown in Ref.~\cite{potts2022dynamicalbackaction}, the measured magnomechanical decay rate exhibits a shift with respect to the theoretical prediction obtained from the Hamiltonian given in Eq.~ \eqref{Eq:1Hamiltonian}. This shift was taken into account by adding to the magnomechanical decay rate a phenomenological correction proportional to $\vert \langle \hat{m} \rangle \vert^2$, which depends on the average number of magnons driven by the microwave tone. Such correction can be attributed to the coupling to additional magnon modes, which are weakly driven by their coupling to the microwave cavity, and to magnon nonlinearities. 

While nonlinearities in magnetic spheres are generally weak, the microwave drive combined with the strong magnon-microwave coupling can make nonlinear effects prominent. This is the case provided that the power of the drive is strong enough to induce an average number of magnons above a certain threshold \cite{lecraw1958ferromagneticresonance}, with implications for magnetoelastic effects \cite{schlomann1960generationofphonons}. Even for drive powers below the nonlinear threshold, magnon nonlinearities can affect the hybrid system dynamics. For instance, in a cavity-magnonic system, the Kittel mode self-Kerr nonlinearity was shown to yield considerable cavity and magnon frequency shifts under moderate driving powers \cite{wang2016magnonkerr}. Experimentally, a phonon frequency shift, as well as mechanical bistability, was reported recently \cite{shen2022mechanical}, which points to the importance of considering such nonlinearities in the description of dynamical backaction effects.

In what follows, we include in the description of dynamical backaction both the coupling to additional magnon modes as well as magnon nonlinearities.

\section{Inclusion of Kerr nonlinearity and coupling to additional magnon modes in the phonon self-energy}
\label{Sec:AdditionalModesSelfEnergy}

To derive the correction term to the self-energy, we consider adding to the Hamiltonian in Eq.~ \eqref{Eq:1Hamiltonian} self- and cross- Kerr nonlinearities, and coupling to $N$ additional magnon modes, each with annihilation operators $\{ \hat{m}_j \}$ and frequencies $\omega_{\rm{j}}$ ($j = 1, ... , N$). The total Hamiltonian is thus
\begin{equation}
\label{App02:Eq01}
\begin{aligned}
\frac{\hat{\mathcal{H}}}{\hbar} &= \omega_{\rm{a}} \hat{a}^\dagger \hat{a} + \omega_{\rm{m}} \hat{m}^\dagger \hat{m} + \Omega_{\rm{b}} \hat{b}^\dagger \hat{b} + \sum_{j = 1}^N \omega_j \hat{m}_j ^\dagger \hat{m}_j \\
& + g_{\rm{am}} \left( \hat{a} \hat{m}^\dagger + \hat{a}^\dagger \hat{m} \right) + \sum_{j = 1}^N g_{\rm{a m_j}} \left( \hat{a} \hat{m}_j^\dagger + \hat{a}^\dagger \hat{m}_j \right) \\
& + g_{\rm{mb}}^0 \hat{m}^\dagger \hat{m} \left( \hat{b}^\dagger + \hat{b} \right) + \sum_{j = 1}^N \hat{m}_j^\dagger \hat{m}_j \left[ g_{\rm{m_j b}}^0 \hat{b}  + (g_{\rm{m_j b}}^{0})^* \hat{b}^\dagger \right] \\
& + K_{\rm{m}} (\hat{m}^\dagger \hat{m})^2 + K_{\rm{cr}} \hat{m}^\dagger \hat{m} \hat{b}^\dagger \hat{b} +  \sum_{j = 1}^N K_{\rm{cr,m}}^{(j)} \hat{m}^\dagger \hat{m} \hat{m}_j^\dagger \hat{m}_j \\
& + i \sqrt{\kappa_e} \epsilon_{\rm{d}}\left(\hat{a} e^{i \omega_{\rm{d}}} - \hat{a}^\dagger e^{-i \omega_{\rm{d}}} \right).
\end{aligned}
\end{equation}
From now on, we will refer to the magnon mode $\hat{m}$ as the Kittel mode, since this is typically the magnon mode that has the strongest coupling to the cavity, while we refer to the magnon modes $\hat{m}_{j}$ as additional magnon modes. We specify such additional magnon modes for the case of a magnetic sphere in Sec.~\ref{Sec:MagnoMechCouplingSphere}. \hl{We furthermore assume a rotating wave approximation for the magnon-microwave coupling, as is done to obtain Eq.~\eqref{Eq:1Hamiltonian}, which eliminates any term of the form $\hat{m}_{(j)} \hat{a}$ and $\hat{m}_{(j)}^\dagger \hat{a}^\dagger$. The rotating wave approximation is also assumed for the magnomechanical coupling, as explained in Sec. \ref{Sec:MagnoMechCoupling}.}

Compared with the Hamiltonian in Eq.~\eqref{Eq:1Hamiltonian}, the above equation includes the following terms: the additional magnon modes; the coupling between these and (i) the microwave mode, each with a coupling rate $g_{\rm{a m_j}}$, and (ii) the phonon mode, each with a coupling rate $g^0_{\rm{m_jb}}$; the self-Kerr term for the Kittel mode; the cross-Kerr term between the Kittel and the phonon modes \cite{wu2021observationofmagnon}; and the cross-Kerr term between the Kittel mode and the other magnon modes. For a sphere, the values of those nonlinear terms depend on the relative orientation between the crystallographic axis [100] of YIG and the bias field \cite{maconald1951ferromagnetic, wang2016magnonkerr, wang2018bistability}, which was not perfectly aligned in the experiment \cite{potts2022dynamicalbackaction} that we use as the case of study. The values for $K_{\rm{m}}$, $K_{\rm{cr}}$, and $K_{\rm{cr, m}}^{(j)}$ that we consider here are effectively smaller than their values in the perfectly aligned case, which we indicate as, e.g., $K_{\rm{m}}^{0}$.  
Provided that the external bias field is aligned with the aforementioned magnetic anisotropy axis of the YIG sphere, the Kittel mode self-Kerr nonlinear coefficient is given by $K_{\rm{m}}^{0} = 13 \hbar K_{\rm{an}} \gamma^2/(16 M_{\rm{s}}^2 V)$, where $K_{\rm{an}} = -610$ J/m$^2$ at room temperature and $V$ is the sphere volume \cite{ stancil2009spinwaves}.  For the experiment in Ref.\,\cite{potts2022dynamicalbackaction} this corresponds to $K_{\rm{m}}^{0}/2 \pi = - 5.15$ nHz. The magnon-phonon and magnon-magnon cross Kerr nonlinear coefficients depend on the overlap between these modes and the Kittel mode. In general, the magnon-magnon cross-Kerr coefficient is around the same order of magnitude as $K_{\rm{m}}^{0}$ \cite{wu2021observationofmagnon}, while the magnon-phonon cross-Kerr coefficient is $\sim  -5$ pHz \cite{shen2022mechanical}. Figure \ref{Fig:App02} shows a schematic of the model, including the different coupling terms.

The values of the magnomechanical couplings depend on the geometry of the magnet, which defines the magnon and phonon mode profiles. In the system under study, the coupling between the Kittel mode and a relevant mechanical mode of a sphere, discussed in Sec. \ref{Sec:MagnoMechCouplingSphere}, is $g^0_{\rm{m b}}/2 \pi= 4.56$ mHz. In Sec. \ref{Sec:MagnoMechCoupling} we discuss in detail the values of $g^0_{\rm{m_jb}}$ for a magnetic sphere. It is important to point out that, due to better mode overlap, in principle, $g^0_{\rm{m_jb}}$ can be comparable to or larger than $g^0_{\rm{mb}}$ for some modes. The coupling between magnons and microwaves depends on the microwave field at the magnet position. For homogeneous fields, only the coupling to the Kittel mode does not vanish. Nevertheless, small inhomogeneities could yield a small microwave-magnon coupling $g_{\rm{a m_j}}$ which would, in turn, drive weakly such magnon modes. For different cavity geometry, such couplings can be strong \cite{rameshti2015magneticspheres, zhang2015cavityquantum, morris2017strong, potts2020strong}, a framework which we do not consider here.

In correspondence with the experiment \cite{potts2022dynamicalbackaction}, we assume that the additional magnon modes are weakly driven via their coupling to the cavity mode, such that we expect a small steady-state amplitude for those modes. Thus we can safely disregard any self- and cross-Kerr nonlinearity of the form $\hat{m}_k^\dagger \hat{m}_k \hat{m}_j^\dagger \hat{m}_j$. The Heisenberg-Langevin equations describing the dynamics of the coupled modes in the rotating frame with the drive frequency are
\begin{equation}
\label{App02:Eq02}
\begin{aligned}
\dot{\hat{a}} &= \left(i \Delta_{\rm{a}} - \frac{\kappa}{2} \right) \hat{a} - i g_{\rm{am}} \hat{m} - i \sum_{j=1}^N g_{\rm{a m_j}} \hat{m}_j \\
&\quad - \sqrt{\kappa_{\rm{i}}} \hat{\xi}_I(t) - \sqrt{\kappa_{\rm{e}}} \epsilon_{\rm{d}}, \\
\dot{\hat{m}} &= \left(i \Delta_{\rm{m}} - \frac{\gamma}{2} \right) \hat{m} - i g_{\rm{am}} \hat{a} - i g_{\rm{mb}}^0 \hat{m} \left( \hat{b}^\dagger + \hat{b} \right)\\
&\quad - i K_{\rm{m}} \hat{m} \left( 1 + 2 \hat{m}^\dagger \hat{m} \right) - i K_{\rm{cr}} \hat{m} \hat{b}^\dagger \hat{b} \\
&\quad - i \sum_{j = 1}^N K_{\rm{cr,m}}^{(j)} \hat{m} \hat{m}_j^\dagger \hat{m}_j + \sqrt{\gamma_{\rm{m}}} \hat{\xi}_{\rm{m}} (t), \\
\dot{\hat{m}}_j & = \left(i \Delta_{\rm{m_j}} - \frac{\gamma_j}{2} \right) \hat{m}_j - i g_{\rm{a m_j}} \hat{a} - i \hat{m}_j\left( g_{\rm{m_j b}}^0 \hat{b} +g_{\rm{m_j b}}^{0,*} \hat{b}^\dagger \right) \\
&\quad - i K_{\rm{cr,m}}^{(j)} \hat{m}_j \hat{m}^\dagger \hat{m} + \sqrt{\gamma_j}\hat{\xi}_{\rm{m}_j}(t), \\
\dot{b} &= -  \left(i \Omega_{\rm{b}} - \frac{\Gamma_{\rm{b}}}{2} \right) \hat{b} - i g_{\rm{mb}}^0 \hat{m}^\dagger \hat{m} - i K_{\rm{cr}} \hat{b} \hat{m}^\dagger \hat{m}\\
& \quad- i \sum_{j = 1}^N  g_{\rm{m_j b}}^0 \hat{m}_j^\dagger \hat{m}_j + \sqrt{\Gamma_{\rm{b}}} \hat{\xi}_{\rm{b}} (t).
\end{aligned}
\end{equation}
In the above equations, $\kappa = \kappa_{\rm{i}} + \kappa_{\rm{e}}$ denotes the total microwave cavity decay rate, which is composed of the intrinsic cavity decay $\kappa_{\rm{i}}$ and the decay into the external port $\kappa_{\rm{e}}$. The additional magnon modes decay rates are indicated by $\gamma_{\rm{m_j}}$ which, for magnetostatic modes of a sphere, have the same value of the Kittel mode decay \cite{klinger2017gilbertdamping}. \hl{The magnon decay $\gamma_{\rm{m}}$ included in our formalism corresponds to the Gilbert damping term included in the Landau-Lifshitz equation to describe magnetic damping.} All parameters appearing in Eq.~\eqref{App02:Eq02} are summarized in Table \ref{Table0}, with the values that will be used throughout this paper. The noise terms denoted by $\hat{\xi}_{\eta}(t)$ ($\eta = {\rm{i}, \rm{e}, \rm{m}, \rm{m_j}, \rm{b} }$) describe thermal (white) noises with correlations
\begin{equation}
\begin{aligned}
\langle \hat{\xi}_{\eta}(t) \hat{\xi}^\dagger_{\eta^\prime}(t^\prime) \rangle &= (n_{{\rm{Th}},\eta}+1) \delta_{\eta \eta^\prime} \delta(t - t^\prime), \\
\langle \hat{\xi}^\dagger_{\eta}(t) \hat{\xi}_{\eta^\prime}(t^\prime) \rangle &= n_{{\rm{Th}},\eta} \delta_{\eta \eta^\prime} \delta(t - t^\prime),
\end{aligned}
\end{equation}
with $n_{{\rm{Th}},\eta} =[{\rm{exp}}(\hbar \omega_\eta /k_{\rm{B}} T) -1]^{-1}$ the number of thermal excitations of mode $\eta$ at a temperature $T$.
\begin{table*}[t!]
\begin{center}
\caption{Parameters of the magnomechanical system appearing in Eq.~\eqref{App02:Eq02}. The values correspond to the experiment in Ref. \cite{potts2022dynamicalbackaction}. The magnon self-Kerr term value corresponds to the case where the bias magnetic field is aligned with the magnetic anisotropy axis. 
\label{Table0}}
\begin{tabular}{ |c|c|c| } 
\hline
\textbf{Parameter} & \textbf{Symbol} & \textbf{Value} \\
\hline
Microwave mode frequency & $\omega_{\rm{a}}$ &  $2 \pi \times 7.11$ GHz \\
\hline
Kittel mode frequency & $\omega_{\rm{m}}$ & $2 \pi \times 7.09$ GHz \\
\hline
Additional magnon modes frequencies & $\omega_{\rm{m_j}}$ & see Sec. II A \\
\hline
Phonon mode frequency & $\Omega_{\rm{b}}$ & $2 \pi \times 12.45$ MHz \\
\hline
Drive frequency & $\omega_{\rm{d}}$ & $2 \pi \times [7.096, 7.099]$ GHz \\
\hline
Microwave intrinsic decay rate & $\kappa_{\rm{i}}$ & $2 \pi \times 2.91$ MHz \\
\hline
Microwave external decay rate & $\kappa_{\rm{e}}$ & $2 \pi \times 3.17$ MHz \\
\hline
Kittel mode decay rate & $\gamma_{\rm{m}}$ & $2 \pi \times 2.55$ MHz \\
\hline
Additional magnon modes decay rate & $\gamma_{\rm{m_j}}$ & $2 \pi \times 2.55$ MHz \\
\hline
Phonon intrinsic decay rate & $\Gamma_{\rm{b}}$ & $2 \pi \times 3.74$ kHz \\
\hline
Kittel mode - Microwave coupling rate & $g_{\rm{am}}$ & $2 \pi \times 9.19$ MHz \\
\hline
Additional magnon modes - microwave coupling rate & $g_{\rm{am_j}}$ & see Sec. \ref{Sec:Evaluation}\\
\hline
Magnomechanical coupling to the Kittel mode & $g_{\rm{mb}}^0$ & $2 \pi \times 4.56$ mHz \\
\hline
Magnomechanical couplings to the $j$-th additional magnon mode & $g_{\rm{m_j b}}^0$ & See section \ref{Sec:MagnoMechCouplingSphere}\\
\hline
Kittel mode self-Kerr nonlinearity & $K_{\rm{m}}^{0}$ & $-2 \pi \times 5.15$ nHz \\
\hline
Magnon cross-Kerr nonlinearity & $K_{\rm{cr, m}}^{(j)}$ & $-2 \pi \times 5.15$ nHz \\
\hline
Magnon-phonon cross-Kerr nonlinearity & $K_{\rm{cr}}$ & $-2 \pi \times 5.4$ pHz \\
\hline
\end{tabular}
\end{center}
\end{table*}

The steady state in a mean-field approximation is obtained by taking the expectation values of the operators in Eqs.~\eqref{App02:Eq02} and ignoring any quantum correlations, i.e.~$\langle \hat{m} \hat{b} \rangle \approx \langle \hat{m} \rangle \langle \hat{b} \rangle$. Since we are assuming that the magnon modes $\{ \hat{m}_j \}$ are weakly coupled to the microwaves, $g_{\rm{a m_j}} g_{\rm{a m_k}} \ll g_{\rm{a m_j}} g_{\rm{a m}}$, we discard any other indirect coupling between the additional magnon modes via the cavity. These approximations yield
\begin{equation}
\label{App02:Eq03}
\begin{aligned}
\langle \hat{b} \rangle &= \frac{i g_{\rm{mb}}^0 \vert \langle \hat{m} \rangle \vert^2}{\mathcal{F}_{\rm{b}} - i K_{\rm{cr}} \vert \langle \hat{m} \rangle \vert^2 } + \frac{i \sum_{j = 1}^N  g_{\rm{m_j b}}^0 \vert \langle \hat{m}_j \rangle \vert^2}{\mathcal{F}_{\rm{b}} - i K_{\rm{cr}} \vert \langle \hat{m} \rangle \vert^2}, \\
\langle \hat{m}_j \rangle &= \frac{i g_{\rm{a m_j}}  \sqrt{\kappa_{\rm{e}}} \epsilon_{\rm{d}} }{\mathcal{F}_{\rm{m_j}} \mathcal{F}_{\rm{a}} + g_{\rm{a m_j}}^2 - i K_{\rm{cr,m}}^{(j)} \vert \langle \hat{m} \rangle \vert^2}  \\
&\quad - \frac{g_{\rm{a m_j}} g_{\rm{a m}} \langle \hat{m} \rangle }{\mathcal{F}_{\rm{m_j}} \mathcal{F}_{\rm{a}} + g_{\rm{a m_j}}^2- i K_{\rm{cr,m}}^{(j)} \vert \langle \hat{m} \rangle \vert^2},
\end{aligned}
\end{equation}
where we have defined
\begin{equation}
\label{App02:Eq04}
\begin{aligned}
\mathcal{F}_{\rm{b}} &=-i \Omega_{\rm{b}} - \frac{\Gamma_{\rm{b}}}{2},  \\
\mathcal{F}_{\rm{ m_j(m)}} &= i \Delta_{\rm{m_j(m)}} - \frac{\gamma_{\rm{m_j(m)}}}{2}, \\
\mathcal{F}_{a} &= i \Delta_{\rm{a}} - \frac{\gamma_{\rm{a}}}{2}.
\end{aligned}
\end{equation}
For $\langle \hat{m}_j \rangle$, we have also discarded the term $\propto g_{\rm{m_j b}}^0$. The steady state of the Kittel mode reads
\begin{equation}
\label{App02:Eq05}
\mathcal{A} \langle \hat{m} \rangle= i  g_{\rm{a m}} \sqrt{\kappa_{\rm{e}}} \epsilon_{\rm{d}}  \mathcal{B}
\end{equation}
where
\begin{equation}
\label{App02:Eq06}
\begin{aligned}
\mathcal{A} &= \mathcal{F}_{\rm{m}} \mathcal{F}_{a} +g_{\rm{am}}^2 - i K_{\rm{m}} \left( 1 + 2 \vert \langle \hat{m} \rangle \vert^2 \right) \\
&\quad -  2 i \mathcal{F}_{a} g_{\rm{mb}}^0  {\rm{Re}}\left[\langle \hat{b} \rangle \right]  + g_{\rm{a m}} ^2 \mathcal{B} - i K_{\rm{cr}} \vert \langle \hat{b} \rangle \vert^2 \\
\mathcal{B} &= 1 - \sum_{j = 1}^N \frac{g_{\rm{a m_j}}^2}{\mathcal{F}_{\rm{m_j}} \mathcal{F}_{\rm{a}} + g_{\rm{a m_j}}^2- i K_{\rm{cr,m}}^{(j)} \vert \langle \hat{m} \rangle \vert^2}.
\end{aligned}
\end{equation}
Equation \eqref{App02:Eq05} is solved numerically. Depending on the drive power and the detuning, the equation can have two bistable solutions. We will focus our analysis on a detuning range lying in between the hybridized Kittel magnon-microwave modes. In the considered range, the magnomechanical decay rate of Eq.~\eqref{Eq:Modifics} changes its sign, and in such a region, the nonlinear equation for $\langle \hat{m} \rangle$ has only one solution. Furthermore, we can discard the terms proportional to $K_{\rm{cr,m}}^{(j)}$, $K_{\rm{cr}}$, and $g_{\rm{mb}}^0$ to obtain the solutions of Eq.~\eqref{App02:Eq05}.

\subsection{Linearized dynamics}

We can now consider the fluctuations around the steady-state values. We write $\hat{o} = \delta \hat{o} + \langle \hat{o} \rangle$, and discard any terms involving more than two fluctuations. The quadratic Hamiltonian describing the dynamics of the fluctuations is given by
\begin{equation}
\label{App02:Eq07}
\begin{aligned}
\frac{\hat{\mathcal{H}}_{\rm{Lin}}}{\hbar} &= -\Delta_{\rm{a}}\delta\hat{a}^{\dagger}\delta\hat{a} + \tilde{\Omega}_{\rm{b}}\delta\hat{b}^{\dagger}\delta\hat{b}- \tilde{\Delta}_{\rm{m}} \delta\hat{m}^{\dagger}\delta\hat{m} \\
&\quad - \sum_{j = 1}^N \tilde{\Delta}_{\rm{m_j}} \delta\hat{m}_j^{\dagger}\delta\hat{m}_j + \frac{\hat{H}_{\rm{Int}}}{\hbar}, 
\end{aligned}
\end{equation}
with the coupling terms included in $\hat{H}_{\rm{Int}}$ given by
\begin{equation}
\label{App02:Eq08}
\begin{aligned}
\frac{\hat{H}_{\rm{Int}}}{\hbar} &= g_{\rm{a m}} \delta\hat{a}^{\dagger}  \delta\hat{m} + G_{\rm{R}} \delta\hat{m}^\dagger \delta \hat{b} + G_{\rm{B}} \delta\hat{m}^\dagger \delta \hat{b}^\dagger \\ 
&\quad + g_{\rm{ms}} \left( \delta \hat{m}^\dagger \right)^2  + \sum_{j = 1 } g_{\rm{a m_j}}  \delta\hat{a}^{\dagger}  \delta\hat{m}_j \\
 &\quad + \sum_{j = 1}^N \left(G_{\rm{R,j}}  \delta\hat{m}_j^\dagger \delta\hat{b}    +  G_{
 \rm{B,j}}   \delta\hat{m}_j \delta\hat{b}   \right) \\
&\quad +\sum_{j = 1}^N  \left(g_{\rm{R,j}}\delta\hat{m}^\dagger \delta\hat{m}_j +  g_{\rm{B,j}}\delta\hat{m}^\dagger \delta\hat{m}^\dagger_j \right ) \\
 &\quad + \rm{H.c.}
\end{aligned}
\end{equation}
The interacting terms appearing in the Hamiltonian of Eq.~\eqref{App02:Eq01} induce frequency shifts for the fluctuations, which are given by 
\begin{equation}
\label{App02:Eq09}
\begin{aligned}
\tilde{\Delta}_{\rm{m(m_j)}} &= \omega_{\rm{d}} - \tilde{\omega}_{\rm{m(m_j)}}, \\
\tilde{\omega}_{\rm{m}} &= \omega_{\rm{m}} +2 g_{\rm{mb}}^0 {\rm{Re}}\left[\langle \hat{b} \rangle \right]+ 4 K_{\rm{m}} \vert \langle \hat{m} \rangle \vert^2 + K_{\rm{cr}}  \vert \langle \hat{b} \rangle \vert^2 \\
&\quad + \sum_{j = 1}^N K_{\rm{cr,m}}^{(j)} \vert \langle \hat{m}_j \rangle \vert^2, \\
\tilde{\omega}_{\rm{m_j}} &= \omega_{\rm{m_j}} +2  {\rm{Re}}\left[ g_{\rm{m_j b}}^0 \langle \hat{b} \rangle \right]+ K_{\rm{cr,m}}^{(j)}  \vert \langle \hat{m} \rangle \vert^2, \\
\tilde{\Omega}_{\rm{b}} &=\Omega_{\rm{b}} + K_{\rm{cr}}  \vert \langle \hat{m} \rangle \vert^2.
\end{aligned}
\end{equation}
The coupling rates between the fluctuations are enhanced and modified with respect to the bare ones. Their expressions are shown in Table \ref{Table01}.
\begin{table}[H]
\begin{center}
\caption{Enhanced couplings appearing in the linearized Hamiltonian in Eq.~\eqref{App02:Eq08} \label{Table01}}
\begin{tabular}{ |c|c| } 
 \hline
 $g_{\rm{mb}}$ & $g_{\rm{mb}}^0  \langle \hat{m} \rangle$ \\ 
 \hline
$G_{\rm{R}}$ & $g_{\rm{mb}}+ K_{\rm{cr}} \langle \hat{m} \rangle \langle \hat{b} \rangle^*$ \\
\hline
$G_{\rm{B}}$ & $g_{\rm{mb}}+ K_{\rm{cr}} \langle \hat{m} \rangle \langle \hat{b} \rangle$ \\ 
 \hline
$G_{\rm{R,j}}$ & $g_{\rm{m_j b}}^0  \langle \hat{m}_j \rangle$ \\
\hline
$G_{\rm{B,j}}$ & $g_{\rm{m_j b}}^0  \langle \hat{m}_j \rangle^*$ \\
\hline
$g_{\rm{ms}}$ & $K_{\rm{m}} \langle \hat{m} \rangle^2$ \\
\hline
$g_{\rm{R,j}}$ & $K_{\rm{cr,m}}^{(j)}\langle \hat{m} \rangle \langle \hat{m}_j \rangle^*$ \\
\hline
$g_{\rm{B,j}}$ & $K_{\rm{cr,m}}^{(j)}\langle \hat{m} \rangle \langle \hat{m}_j \rangle$ \\
\hline
\end{tabular}
\end{center}
\end{table}

\subsection{Calculation of the phonon self-energy}

The phonon self-energy is obtained by solving the linear Heisenberg-Langevin equations describing the coupled dynamics of the fluctuations for the phonon operator. To compute the effects of backaction in the response of the phonon mode to noise, we consider the Fourier transformed operators defined by
\begin{equation}
\hat{o}(t)= \int d\omega e^{- i \omega t} \hat{o}[\omega],
\end{equation}
where $\hat{o} = \delta \hat{a}^{(\dagger)}, \delta \hat{m}^{(\dagger)},\delta \hat{m}_j^{(\dagger)}, \delta \hat{b}^{(\dagger)}$.
We skip the algebraic steps, but outline the main differences with respect to the results in Refs.\,\cite{potts2020magnon, potts2021dynamical}. After writing the cavity operator in terms of the magnon operators, we obtain the following equation for the additional magnon modes:
\begin{equation}
\label{App02:Eq10}
\begin{aligned}
\Xi_j[\omega]^{-1} \delta \hat{m}_j[\omega] &= -i \left( g_{\rm{R,j}}^* -i g_{\rm{a m_j}} g_{\rm{a m}} \chi_{\rm{a}}[\omega]  \right) \delta \hat{m}[\omega] \\
& \quad - i g_{\rm{B,j}} \delta \hat{m}^\dagger[\omega] - i G_{\rm{R,j}} \delta \hat{b}[\omega] - i G_{\rm{B,j}}^* \delta \hat{b}^\dagger[\omega]\\
&\quad + g_{\rm{a m_j}} \chi_{\rm{a}}[\omega]  \sum_{k\neq j} g_{\rm{a m_k}} \delta \hat{m}_k [\omega] + \hat{\tilde{\xi}}_{\rm{m_j}}[\omega],
\end{aligned}
\end{equation}
where $ \hat{\tilde{\xi}}_{\rm{m_j}}$ represents the noise term modified by the interaction with the cavity, and we have defined the effective magnon susceptibility in correspondence with the previous case in Eq.~\eqref{Eq:XiKit},
\begin{equation}
\label{App02:Eq11}
\Xi_j[\omega]^{-1} = \chi_{\rm{m_j}}^{-1}[\omega] +  g_{\rm{a m_j}}^2 \chi_{\rm{a}}[\omega]. 
\end{equation}
We notice that the first term on the right-hand side of Eq.~\eqref{App02:Eq10} includes the indirect coupling between the $j$-th magnon mode and the Kittel mode via the cavity. A similar term related to the coupling between the additional magnon modes appears in the last line of Eq.~\eqref{App02:Eq10}, and since $g_{\rm{a m_j}} \ll g_{\rm{a m}}$, we discard these contributions.

After using Eq.~\eqref{App02:Eq10} to eliminate the additional magnon modes in favor of the Kittel mode and the phonon fluctuations, we obtain the following set of coupled equations
\begin{equation}
\label{App02:Eq12}
\begin{aligned}
\Xi_{\rm{m}}^{-1} [\omega]\delta \hat{m}[\omega] &= \eta_{\rm{m}}[\omega] \hat{m}[\omega]- i \Lambda_{\rm{m}}[\omega] \hat{m}^\dagger[\omega] \\
&\quad - i G_{\rm{m,R}} [\omega] \delta \hat{b}[\omega]  \\
&\quad - i G_{\rm{m,B}}[\omega] \delta \hat{b}^{\dagger}[\omega]  + \hat{\tilde{\xi}}_{\rm{m}}[\omega], \\
\left [\chi_{\rm{b}}^{-1}[\omega] - i \sum_{j = 1}^N \sigma_j[\omega] \right] \delta \hat{b}[\omega] &= - i \tilde{G}_{\rm{b,R}}^*[-\omega] \delta \hat{m}[\omega] \\
&\quad  - i \tilde{G}_{\rm{b,B}}[\omega] \delta \hat{m}^\dagger [\omega]+ \hat{\tilde{\xi}}_{\rm{b}}[\omega] \\
 &\quad+ i \left( \sum_{j = 1}^N \lambda_j[\omega] \right) \delta \hat{b}^\dagger [\omega].
\end{aligned}
\end{equation}
We have included all the noise terms in $\hat{\tilde{\xi}}_{\rm{m},\rm{b}}[\omega]$; all other functions appearing in the equations below are defined in the following. The coupling to the additional magnon modes has the following effects: the introduction of a self-energy term on the phonon mode, the modification of the coupling between the Kittel mode and the phonon mode, and a modification of the Kittel mode susceptibility and squeezing. We briefly comment on each of these effects. 

At this intermediate step, the phonon susceptibility is modified by the self-energy term
\begin{equation}
\label{App02:Eq13}
\sum_{j = 1}^N \sigma_j[\omega] = i \sum_{j = 1}^N \vert g_{\rm{m_j b}}^0 \langle \hat{m}_j \rangle \vert^2 \left(\Xi_j[\omega]- \Xi_j^*[\omega]\right),
\end{equation}
which is defined in analogy with the self-energy term derived in Refs. \cite{potts2020magnon, potts2021dynamical} and given in Eq.~\eqref{Eqn:06}. Such a term represents the direct dynamical backaction of the coupling between the phonon mode and the additional magnon modes.

The additional magnon modes modify the couplings between the Kittel mode and the phonon mode. In fact, the effective coupling constants appearing in Eqs.~\eqref{App02:Eq12}, \hl{whose explicit forms are given in Appendix \ref{AppA}, include two types of modifications. The first is an indirect coupling between the additional magnon modes and the Kittel mode via the cavity. The second is a term proportional to $g_{\rm{B(R),j}}$, which in turn (see Table \ref{Table01}) is due to the magnon cross-Kerr nonlinearity, $\hat{m}_j^\dagger \hat{m}_j \hat{m}^\dagger \hat{m}$ in the Hamiltonian of Eq.~\eqref{App02:Eq01}. The relevance of these corrections for a given drive frequency is determined by the susceptibilities of the additional magnon modes.}

The Kittel mode squeezing term $\Lambda_{\rm{m}}[\omega]$ reads
\begin{equation}
\label{Eq:Lambdam}
\begin{aligned}
\Lambda_{\rm{m}}[\omega]&= 2 g_{\rm{ms}}  \\ & \quad - g_{\rm{am}} (\chi_{\rm{a}}[\omega] + \chi_{\rm{a}}^*[-\omega]) \sum_{j} g_{\rm{B,j}} g_{\rm{a m_j}} \Xi_j [\omega],
\end{aligned}
\end{equation}
where the first term is due to the self-Kerr nonlinearity, while the second term is a combination of the magnon cross-Kerr nonlinearity with the indirect coupling between the magnon modes via the microwave cavity. The Kittel mode susceptibility is also modified by the term $\eta_{\rm{m}}[\omega]$, which is given by
\begin{equation}
\label{Eq:KittelSuscept}
\begin{aligned}
\eta_{\rm{m}}[\omega] &= g_{\rm{am}}^2 \chi_{\rm{a}}^2[\omega] \sum_j g^2_{\rm{am_j}} \Xi_j[\omega] \\
&+ 2 i  g_{\rm{am}} \chi_{\rm{a}}[\omega] \sum_{j} {\rm{Re}}[g_{\rm{R,j}}] g_{\rm{am_j}} \Xi_j[\omega] \\
&- \sum_j \left( \vert g_{\rm{R,j}} \vert^2 \Xi_j [\omega] - \vert g_{\rm{B,j}} \vert^2 \Xi_j^*[-\omega] \right).
\end{aligned}
\end{equation}
The three terms in Eq.\,\eqref{Eq:KittelSuscept} describe the effects of a two-mode squeezing between the Kittel mode and each of the additional magnon modes. The first term is related to the indirect coupling of the modes via the microwave cavity while the last term is the direct two mode squeezing induced by the magnon cross-Kerr nonlinearity. The second term is a combination of both effects.

From Eqs.~\eqref{App02:Eq12}, we eliminate the Kittel mode operator and obtain an equation for the phonon mode operator,
\begin{equation}
\label{App02:Eq14}
\begin{aligned}
\left [\chi_{\rm{b}}^{-1}[\omega] - i \Sigma_{\rm{Tot}} [\omega] \right] \delta \hat{b}[\omega] = i \Lambda_{\rm{b}}[\omega] \delta \hat{b}^\dagger [\omega] + \hat{\Upsilon}_{\rm{b}}[\omega],
\end{aligned}
\end{equation}
where $\hat{\Upsilon}_{\rm{b}}[\omega]$ includes all the noise terms driving the phonon fluctuations, $ \Lambda_{\rm{b}}[\omega]$ describes phonon squeezing, and $ \Sigma_{\rm{Tot}} [\omega]$ is the total self-energy. We focus only on the self-energy term, which is given by
\begin{equation}
\label{App02:Eq15}
\Sigma_{\rm{Tot}} [\omega] = \Sigma_{\rm{m}}[\omega] + \sum_{j = 1}^N \sigma_j[\omega],
\end{equation}
where the contribution of Kittel mode to the self-energy is given by
\begin{equation}
\label{App02:Eq16}
\begin{aligned}
\Sigma_{\rm{m}}[\omega] &= i \Big( \tilde{G}_{\rm{b, R}}^*[-\omega] \tilde{G}_{\rm{m,R}}[\omega] \tilde{\Xi}_{\rm{m}} [\omega]  \\
&\quad \quad -  \tilde{G}_{\rm{b, B}}[\omega] \tilde{G}_{\rm{m, B}}^*[-\omega] \tilde{\Xi}_{\rm{m}}^* [-\omega] \Big).
\end{aligned}
\end{equation}
\hl{The modified magnomechanical couplings $\tilde{G}_{\rm{m,R(B)}}[\omega]$ are given in Appendix \ref{AppA}. The Kittel mode susceptibility
\begin{equation}
\tilde{\Xi}_{\rm{m}}^{-1}[\omega] = \Xi^{-1}_{\rm{m}}[\omega] - \eta_{\rm{m}}[\omega] - \frac{\Lambda_{\rm{m}}[\omega] \Lambda^*_{\rm{m}}[-\omega]}{\Xi_{\rm{m}} ^{*,-1}[-\omega]-\eta_{\rm{m}} ^*[-\omega]},
\end{equation}
which includes both the Kittel magnon squeezing in $\Lambda_{m}[\omega]$ and the effects of two-mode squeezing interactions with the additional magnon modes in $\eta_{\rm{m}}[\omega]$.}

\subsection{Magnomechanical decay rate corrections}

We turn our attention now to the effects on the magnomechanical decay rate, in connection with the observed shift reported in Ref.\,\cite{potts2022dynamicalbackaction}. The total change in the phonon linewidth is given by
\begin{equation}
\label{App02:Eq17}
\begin{aligned}
\Gamma_{\rm{mag}}[\omega] &= 2 {\rm{Im}}[\Sigma_{\rm{Tot}} [\omega]] = 2 {\rm{Im}}[\Sigma_{\rm{m}} [\omega]] + \sum_{j = 1}^N \Gamma_j [\omega] \\
\Gamma_j [\omega] &= 2 {\rm{Im}}[\sigma_j[\omega]].
\end{aligned}
\end{equation}
The contributions of self- and cross-Kerr nonlinear terms are included in the above self-energy. The corresponding frequency shift is different than the static one, observed in Ref.\, \cite{shen2022mechanical}, which has already been included in the modified phonon frequency $\tilde{\Omega}_b$ defined in Eqs.~\eqref{App02:Eq09}. The self-Kerr nonlinearity changes the behavior of the magnomechanical decay rate with the detuning. This is due to three main factors: the modification of the steady-state number of magnons $\vert \langle \hat{m} \rangle \vert^2$, the induced static magnon frequency shift, given in Eq.\,\eqref{App02:Eq09}, and the generation of squeezing in the magnon fluctuations. This has a consequence for both dynamical backaction evasion, as we will show, and for backaction cooling. The latter has been recently reported in a system where a mechanical oscillator is parametrically coupled to a nonlinear cavity \cite{zoepfl2022kerrenhanced}.  The magnomechanical frequency shift is given by
\begin{equation}
\delta \Omega_{\rm{b}} [\omega] = - \rm{Re}\left[\Sigma_{\rm{Tot}} [\omega] \right],
\end{equation}
which has a decomposition similar to the one shown in Eq. \eqref{App02:Eq17}.

The relevance of the different corrections due to the additional magnon modes depend on their frequencies with respect to the drive and their coupling rates to the cavity. Recalling that $G_{j} = g_{\rm{m_j b}}^0  \langle \hat{m}_j \rangle$, we use the steady state from Eq.\,\eqref{App02:Eq03}. For the experimental parameters in consideration, given in Table~\ref{Table0}, due to the strong coupling between the Kittel mode and the cavity, $\sqrt{\kappa_{\rm{e}}} \epsilon_{\rm{d}} \ll g_{\rm{a m}} \rm{Re}[\langle \hat{m} \rangle]$ and the steady state of the weakly driven Walker modes can be approximately written as
\begin{equation}
 \label{App02:Eq18}
\vert \langle \hat{m}_j \rangle \vert^2 = \frac{ g_{\rm{a m_j}}^2 g_{\rm{a m}}^2 \vert \langle \hat{m} \rangle \vert^2 }{\vert \mathcal{F}_{\rm{m_j}} \mathcal{F}_{\rm{a}} + g_{\rm{a m_j}}^2 - i K_{\rm{cr,m}}^{(j)} \vert \langle \hat{m} \rangle \vert^2 \vert^2}.
 \end{equation}
Within these approximations, the contribution of such Walker modes to the magnomechanical linewidth is given by
 \begin{equation}
 \label{App02:Eq19}
 \begin{aligned}
 \Gamma_j [\omega] &= 2 g_{\rm{a m}}^2  \vert \langle \hat{m} \rangle \vert^2  \frac{\vert g_{\rm{m_j b}}^0 \vert^2 g_{\rm{a m_j}}^2 }{\vert \mathcal{F}_{\rm{m_j}} \mathcal{F}_{\rm{a}} + g_{\rm{a m_j}}^2 - i K_{\rm{cr,m}}^{(j)} \vert \langle \hat{m} \rangle \vert^2 \vert^2} \\
 &\quad \quad \times \rm{Re}\left( \Xi_j[\omega]- \Xi_j^*[\omega] \right),
 \end{aligned}
 \end{equation}
which for small $K_{\rm{cr,m}}^{(j)}$ gives a contribution to the magnomechanical decay rate which is proportional to $\vert \langle \hat{m} \rangle \vert^2$. The contribution to the magnomechanical decay rate also depends on the detuning between the drive and the magnon frequency.

While $\Gamma_j[\omega]$ quantifies the direct effect of the coupling between the phonon mode and the additional magnon modes, there are also \textit{indirect} effects due to the coupling between the additional magnon modes and the Kittel mode via the microwave cavity. Those are included in the term $2 {\rm{Im}}[\Sigma_{\rm{m}} [\omega]]$ via the modified coupling rates $\tilde{G}_{\rm{m(b), R(B)}}[\omega]$ and the modified Kittel mode susceptibility $\tilde{\Xi}_{\rm{m}}[\omega]$. In general, the corrections included in those terms are proportional to $\vert g_{\rm{ms}} \vert ^2$, to $\vert G_{R (B),j}\vert^2$ or to $\vert g_{R(B),j} \vert^2$. Following the same procedure outlined above, it is possible to show that those are all proportional to the steady-state Kittel mode occupation. Their frequency-dependent coefficients have a more complicated form due to the explicit dependence of the modified couplings and susceptibilities on frequency. We can describe all the corrections by the expression
\begin{equation}
\Gamma_{\rm{mag}}[\omega] = \Gamma_{\rm{mag}}^{0}[\omega] + \alpha[\omega] \vert \langle \hat{m} \rangle \vert^2,
\end{equation}
where $\Gamma_{\rm{mag}}^0[\omega]$ is given in Eq. \eqref{Eqn:06}, without including additional magnon modes and nonlinearities. Such a correction was used phenomenologically in Ref.\, \cite{potts2022dynamicalbackaction} to explain the observed discrepancies of the experimental results with $\Gamma_{\rm{mag}}^0[\omega]$.

\section{Parametric magnomechanical coupling for a magnetic sphere}
\label{Sec:MagnoMechCoupling}

In the previous section, we obtained the modifications of the phonon self-energy due to the coupling between the phonon mode and additional magnon modes. Such contributions depend on the relative strength of the magnomechanical coupling to the additional magnon modes with respect to the coupling to the Walker mode, which in turn depends on the geometry of the magnet. Before specifying the latter, we briefly review the main points of the derivation of the magnomechanical coupling Hamiltonian, done in detail in the literature \cite{zhang2016cavity,Gonzalez_2020_Theory, engelhardt2022optimalbroadband, fan2023stationaryoptomagnonics}, starting from the magnetoelastic energy \cite{kittel1949physicaltheory, landau1984electrodynamics}
\begin{equation}
\label{Eq:MagnetoElastic}
\begin{aligned}
E_{\rm{ME}} &= \frac{B_1}{M_{\rm{S}}^2} \int d^3 r \, \left(M_x^2 \varepsilon_{xx} + M_y^2 \varepsilon_{yy} + M_z^2 \varepsilon_{zz} \right) \\
&\quad + \frac{2 B_2}{M_S} \int d^3 r \, \Big( M_x M_y \varepsilon_{xy} + M_{y} M_z \varepsilon_{yz} \\
&\quad \quad \quad \quad \quad \quad \quad \quad + M_x M_z \varepsilon_{xz}  \Big),
\end{aligned}
\end{equation}
where $M_{x,y,z}$ are the magnetization components, $M_S$ is the saturation magnetization, and
\begin{equation}
\varepsilon_{ij} = \left( \partial_i u_j + \partial_j u_i \right)/2
\end{equation}
is the linear strain tensor with $\bm{u}(\bm{r},t)$ the displacement. The magnetoelastic coefficients $B_1$ and $B_2$ are material- and temperature-dependent constants. For YIG at room temperature, $M_S = 140$ $\rm{kA/m}$ \cite{stancil2009spinwaves}, $B_1 = 3.48 \times 10^{5}$ $\rm{J/m^3}$ and $B_1 = 6.4 \times 10^{5}$ $\rm{J/m^3}$ \cite{zhang2016cavity}. The magnomechanical Hamiltonian is then obtained by quantizing the magnetization and displacement fields. \hl{ The procedure yields both resonant and parametric interactions. The first is relevant for nanometer-sized magnets \cite{Gonzalez_2020_Theory}, while the second is relevant when the magnon mode is not resonant with the phonon mode, which is typically the case for larger magnets. Details are shown in Appendix \ref{AppB}.}

Here, we focus \hl{on} the parametric phonon-magnon coupling, which is given by the Hamiltonian
\begin{equation}
\label{Eq:GenParHamilt}
\begin{aligned}
\hat{\mathcal{H}}_{\rm{mb}}/\hbar &=\sum_{\{j\neq k\}, \alpha} \left[g_{\rm{m_k  m_j  b_\alpha}}^{0} \hat{m}_k^\dagger \hat{m}_j \hat{b}_\alpha + \tilde{g}_{\rm{m_k  m_j  b_\alpha}}^{0} \hat{m}_k^\dagger \hat{m}_j \hat{b}^\dagger_\alpha  \right]  \\
&\quad + \sum_{j, \alpha} g_{\rm{m_j  b_\alpha}}^{0} \hat{m}_j ^\dagger \hat{m}_j \hat{b}_\alpha + \rm{H.c.},
\end{aligned}
\end{equation}
where $\{j\neq k\}$ indicates that the sum is over all $j$'s not equal to $k$ and without repeating combinations. In the above equation we have separated the coupling terms between one magnon mode and one phonon mode and the terms involving two different magnon modes and a phonon mode. \hl{The explicit forms of the couplings are given in Ref. \cite{Gonzalez_2020_Theory} and in Appendix \ref{AppB} in Eq.~\eqref{Eq:Couplings}. They depend on overlap integrals involving the magnon mode functions as well as the derivatives of the displacement field.}

Focusing now on the coupling to a specific phonon mode, the magnomechanical Hamiltonian is given by
\begin{equation}
\begin{aligned}
\frac{\hat{\mathcal{H}}_{\rm{mb}}}{\hbar} &= \omega_{\rm{m}} \hat{m}^\dagger \hat{m}+ \sum_{j} \omega_{j} \hat{m}_{j}^\dagger \hat{m}_j + \Omega_{\rm{b}} \hat{b}^\dagger \hat{b} \\
&\quad + \frac{\hat{\mathcal{H}}_{\rm{mb},I}}{\hbar},
\end{aligned}
\end{equation}
where the coupling terms are
\begin{equation}
\label{Eq:MagnomechHi}
\begin{aligned}
\frac{\hat{\mathcal{H}}_{\rm{mb},I}}{\hbar} &= g_{ \rm{mb}}^0 \hat{m}^\dagger \hat{m} \hat{b} + \sum_{j} g_{\rm{m_j b}}^0 \hat{m}_j^\dagger \hat{m}_j \hat{b} \\
&\quad + \hat{m}^\dagger \sum_j g_{\rm{m m_j b }}^0 \hat{m}_{j} \hat{b} + \sum_{j\neq k} g_{\rm{m_k m_j b }}^0 \hat{m}_k^\dagger \hat{m}_{j} \hat{b}  \\
&\quad + {\rm{H.c.}}
\end{aligned}
\end{equation}
In the above equations, we have separated the terms of the Kittel mode, which from now on we do not label, while the other Walker modes are labeled by the index $j$. As shown in Appendix \ref{AppD}, we can absorb the phase of the coupling to the Kittel mode in the phonon field, such that $ g_{ \rm{mb}}^0$ is real after such a transformation.

\subsection{Magnomechanical coupling rates for a sphere}
\label{Sec:MagnoMechCouplingSphere}

\begin{figure}
\includegraphics[width = \columnwidth]{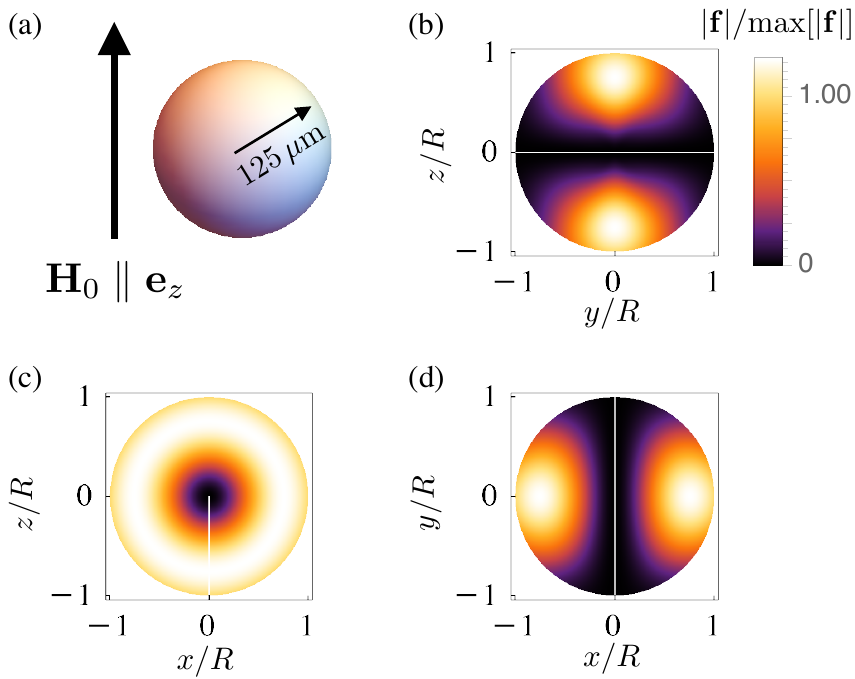}
\caption{Profile of the $S_{122}$ mode of a sphere. (a) The bias field $\bm{H}_0$ is parallel to the $\bm{e}_z$ direction, and we consider a sphere made of YIG with a radius $R=125$ $\mu$m. (b-d) Mode profile $\vert \bm{f}(\bm{r}) \vert$ for the spherical mode $S_{122}$ in the (b)$yz$, (c) $xz$, and (d) $zy$ planes.}
\label{Fig:PhononMode}
\end{figure}

The magnomechanical coupling rates depend on the specific geometry of the sample and the direction of the applied magnetic field, which defines the mode functions and their overlap. We \hl{evaluate now the coupling for} a YIG sphere, corresponding to the experimental configuration of Ref. \cite{potts2022dynamicalbackaction}.

A sphere supports magnetostatic modes called Walker modes \cite{walker1958resonant,fletcher1959ferrimagnetic, roschmann1977propertiesof}, which have frequencies that can be tuned by the value of the external bias field. \hl{The Walker modes are labeled by three indices $\{ l m \nu \}$, with $l\ge 1$ and $\vert m \vert \le l$. For $m>0$, there are $(l-\vert m \vert)/2$ (rounded down) modes labeled with the third index $\nu$, while for $m<0$, there are $1+ (l-\vert m \vert)/2$ (rounded down) modes.  For the phonon modes, considering an unpinned sphere under stress-free boundary conditions, there are two families of mechanical modes: torsional ($\rm{T}$) and spherical ($\rm{S}$). Each phonon mode is labeled by three indices $\{\nu l m\}$, where $l$ and $m$ are polar and azimuthal indices while $\nu$ is a radial index. In correspondence with the experiment \cite{potts2022dynamicalbackaction}, we consider in the following the spherical mode $S_{122}$, shown in Fig~\ref{Fig:PhononMode}. Further details on the magnon and spherical phonon modes of a magnetic sphere are given in Appendix \ref{AppC}.

Even though the magnon and phonon mode functions are given in terms of well-known special functions, the coupling constant (see Eq.~\eqref{Eq:Couplings}) involves a nontrivial combination of derivatives of those. Furthermore, the derivation of the Walker modes involves a transformation to a nonorthogonal coordinate system, which is not easily invertible. While this is not a problem when computing the coupling to the Kittel mode, which is uniform, the exact expressions for coupling rates are not elucidating.} Differently from other parametrically coupled systems, it is hard to infer, for example, selection rules. We, therefore, compute the overlap integrals numerically and evaluate how the couplings $g_{\rm{m_j b}}^{0}$ compare with the coupling to the Kittel mode $g_{\rm{m b}}^{0}$. It is also important to notice that the magnomechanical couplings depend on both the intensity of the bias magnetic field and its direction. In fact, the coupling to the Kittel mode can even vanish for specific relative orientation of the magnetic field \cite{zhang2016cavity}. We consider the case of a fixed bias field at a direction that maximizes the coupling between the Kittel and the $S_{122}$ modes, as depicted in Fig. \ref{Fig:PhononMode}.

\begin{figure}[H]
\includegraphics[width = \columnwidth]{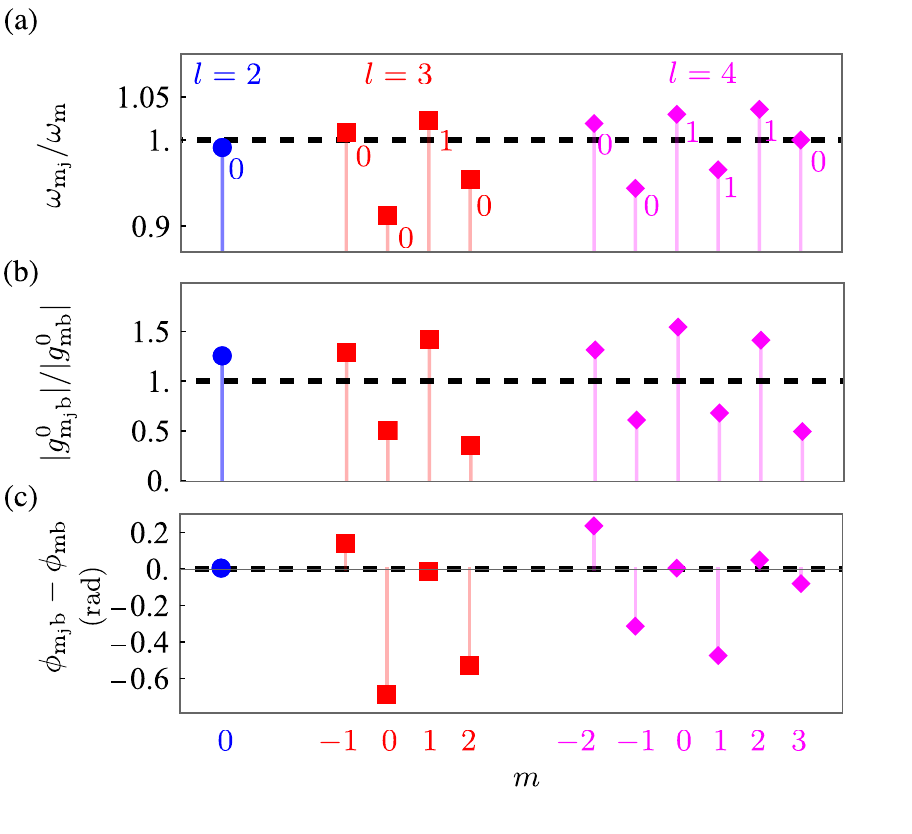}
\caption{(a) Frequency of the Walker modes $\omega_{\rm{m_j b}}$ in units of the Kittel mode frequency $\omega_{\rm{m b}}$. The labels by each point indicate the radial magnon mode label $\nu$; (b) Absolute value of the magnomechanical coupling between the Walker modes and the $S_{122}$ mode in units of the coupling to the Kittel mode $g_{\rm{m b}}^{(0)}$; (c) Phase of the magnomechanical coupling with respect to the phase of the Kittel mode magnomechanical coupling $\phi_{\rm{m_j b}}-\phi_{\rm{m b}}$ in radians. Results for a sphere of radius $R = 125$ $\mu \rm{m}$. The dashed line is the reference value (for the Kittel mode) for each quantity.}
\label{Fig:Couplings}
\end{figure}

Figure \ref{Fig:Couplings} shows the frequencies $\omega_{\rm{m}_j}$ of the Walker modes, the ratios $\vert g_{\rm{m_j b}}^{0} \vert /\vert  g_{\rm{m b}}^{0} \vert$ between the magnomechanical coupling rate to the Walker mode $(l m \nu)$ and to the Kittel mode, and $\phi_{\rm{j}} -\phi$, the relative phase between $g_{\rm{m_j b}}^{0}$ and $g_{\rm{m b}}^{0}$. Results are shown for $l$ up to 4 and for Walker modes lying in a frequency range close to the Kittel mode. Due to better mode overlap, some higher order Walker modes, for example, the $(200)$, couple strongly with the phonon mode in comparison with the coupling to the Kittel mode. In the theoretical analysis of Sec. I, we have not included in the magnomechanical Hamiltonian in Eq.\,\eqref{App02:Eq01} the last two terms of Eq.\,\eqref{Hamiltonian0}. Those describe scattering processes between different magnon modes via the phonon mode. For the considered case, $\vert g_{\rm{m m_j b}}^0 \vert ,\vert g_{\rm{m_k m_j b}}^0 \vert \ll \vert g_{\rm{m b}}^0 \vert, \vert g_{\rm{m_j b}}^0 \vert$, and those processes can be safely discarded. Nevertheless, it is possible that for some relative orientation between the magnon modes and the phonon mode, set by the external bias field, those processes can have a stronger coupling rate.

\section{Evaluation of the model for the phonon self-energy on dynamical backaction evasion}
\label{Sec:Evaluation}

The self-energy obtained in Eq. \eqref{App02:Eq17} includes contributions due to the Kerr nonlinearity and to the couplings to higher-order Walker modes. We focus our analysis now on the effect of such contributions to the magnomechanical decay for detunings in the vicinity of the backaction evasion point.

\begin{figure}[H]
\includegraphics[width = 0.9\columnwidth]{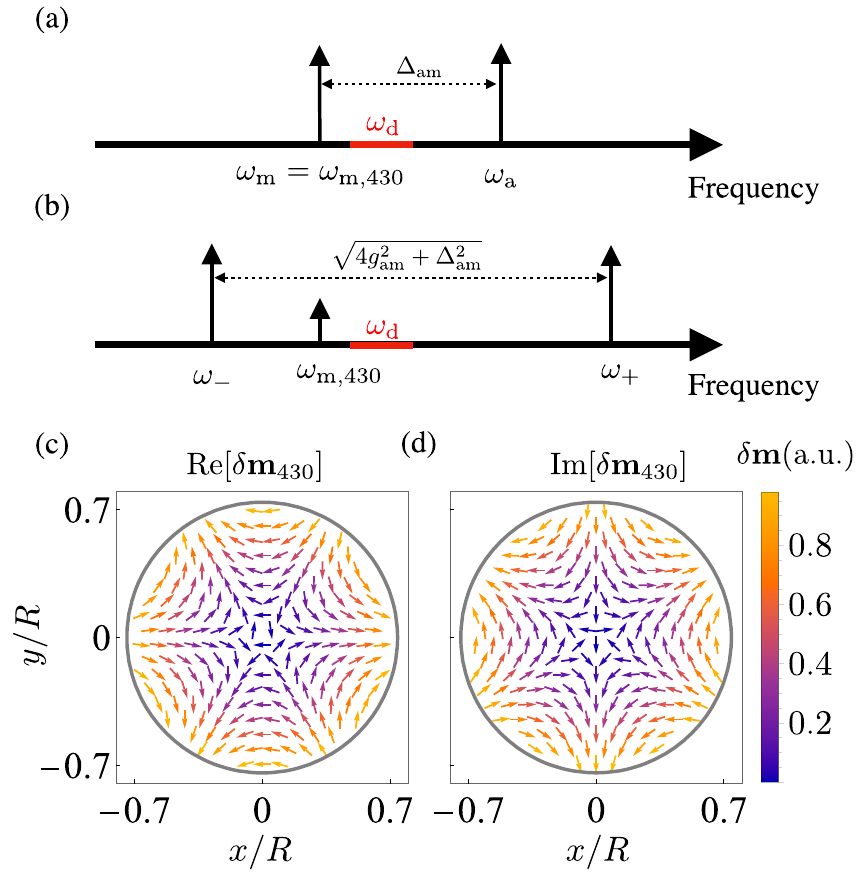}
\caption{Frequency configuration of the magnomechanical system in consideration. (a) The microwave cavity frequency $\omega_{\rm{a}}$ is higher than the Kittel mode frequency, which is degenerate with the $(4,3,0)$ Walker mode. The red frequency range corresponds to the microwave drive considered here. (b) Due to strong coupling, the Kittel mode and the microwave mode hybridize, forming the two modes $\omega_{\pm}$. (c) Real part and (d) imaginary part of the transverse magnetization of the Walker mode $(4,3,0)$. The profiles were evaluated at $z/R = \cos{\pi/4}$, and for better visualization, the vectors were normalized to ${\rm{max}}[\sqrt{\rm{Re}[\delta m_x]^2 + \rm{Re}[\delta m_y]^2}]$.}
\label{Fig:MagnonMode}
\end{figure}

In correspondence with the experiment \cite{potts2022dynamicalbackaction}, we consider that the microwave drive frequency is varied between $\omega_{\rm{d},-}$ and $\omega_{\rm{d},+}$ inside the frequency range $\{\omega_-, \omega_+ \}$ between the hybrid mode frequencies, given by Eq. \eqref{Eq:HybridModeFrequencies}. The Walker modes contributing appreciably to the phonon self-energy lie between $(\omega_{d,-}-\Omega_{\rm{b}})$ and  $(\omega_{d,+}+\Omega_{\rm{b}})$. Since the system is in the resolved sideband regime, any modes outside this frequency range would not allow efficient scattering of phonons, and thus can be neglected. The first condition corresponds to the lower drive frequency corresponding to the blue sideband of a magnon mode, while the second corresponds to driving the red sideband of a magnon mode. For the parameters summarized in Table \ref{Table0}, only the mode $(4,3,0)$ is in this frequency range. In fact, the $(4,3,0)$ mode is degenerate with the Kittel mode. The frequency configuration is shown in Figs. \ref{Fig:MagnonMode}(a,b), and the mode profile of the Walker mode $(4,3,0)$ is shown in Figs. \ref{Fig:MagnonMode} (c,d).

To address the effects of nonlinearities and coupling to the higher order Walker mode, we define the dimensionless parameters $\eta_c = g_{\rm{a m_j}}/g_{\rm{a m}}$ and $\eta_K = K_{\rm{m}}/K_{\rm{m}}^{0}$. The first parameter quantifies the strength of the coupling between the Walker modes and the cavity compared to the coupling between the Kittel mode and the cavity. The second parameter quantifies the strength of the self-Kerr nonlinearity compared to the value shown in Table \ref{Table0} $K_{\rm{m}}^{0}=- 2 \pi \times 5.15$ nHz. We call here $\eta_K$ the dimensionless Kittel magnon self-Kerr nonlinearity. This parameter depends on the alignment between the anisotropy axis of the magnet with the external magnetic field, which has not been taken into account in Ref. \cite{potts2022dynamicalbackaction}.

\begin{figure}[t]
\includegraphics[width = \columnwidth]{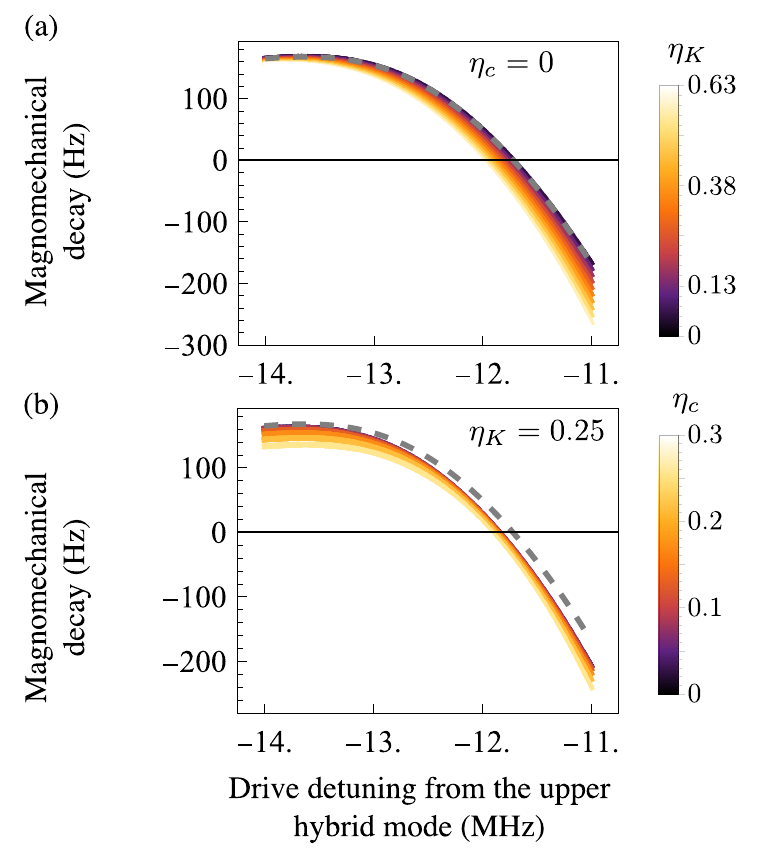}
\caption{Magnomechanical decay rate $\Gamma_{\rm{mag}}[\Omega_b]$ including the contribution of the Walker mode $(4,3,0)$ as a function of the detuning from the upper hybrid mode for (a) $\eta_c = 0$ (without microwave coupling to the additional Walker mode) and for several values of $\eta_K$ (dimensionless Kittel magnon self-Kerr nonlinearity) and (b) for $\eta_K = 0.25$ and for several values of $\eta_c$. The dashed line is the prediction from the self-energy \eqref{Eqn:06} derived in Ref. \cite{potts2020magnon}. The magnomechanical coupling to the $(4,3,0)$ Walker mode corresponds to that shown in Fig.\,\ref{Fig:Couplings}. The driving power is $15$ mW. Parameters in correspondence with the experiment \cite{potts2022dynamicalbackaction}, given in Table \ref{Table0}.}
\label{Fig:DecaysOne}
\end{figure}

Figure \ref{Fig:DecaysOne}(a) shows the magnomechanical decay for $\eta_c = 0$ (the additional Walker mode is not driven by the microwaves) and for several values of $\eta_K$, and Fig.~\ref{Fig:DecaysOne}(b) the magnomechanical decay for several values of $\eta_c$ at a fixed $\eta_K = 0.2$. The self-Kerr nonlinearity of the Kittel mode changes the slope of the magnomechanical decay as a function of the detuning. For a fixed Kerr nonlinearity, the additional magnon mode shifts down the magnomechanical decay; that is, the weakly driven Walker mode adds energy to the vibrational mode, yielding a negative contribution to the decay rate.

\begin{figure}
\includegraphics[width = \columnwidth]{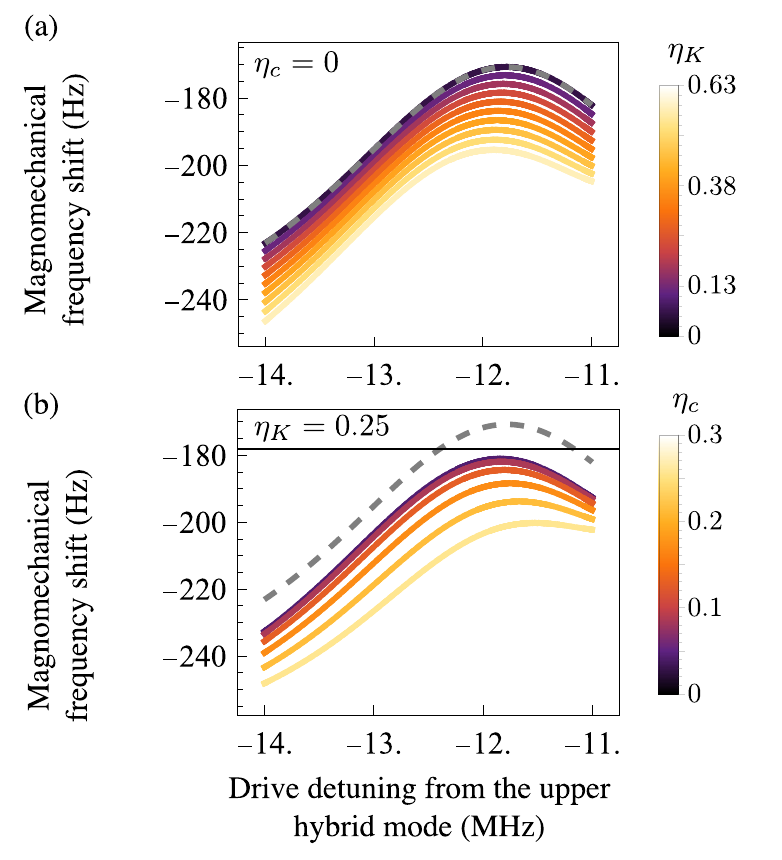}
\caption{Magnomechanical frequency shift $\delta \Omega_{\rm{b}}$ including the contribution of the Walker mode $(4,3,0)$ as a function of the detuning from the upper hybrid mode for (a) $\eta_c = 0$ (without microwave coupling to the additional Walker mode) and for several values of $\eta_K$ (dimensionless magnon self-Kerr nonlinearity); and (b) for $\eta_K = 0.25$ and for several values of $\eta_c$. The dashed line is the prediction from the self-energy Eq .~\eqref{Eqn:06} derived in Ref.~\cite{potts2020magnon}. The magnomechanical coupling to the $(4,3,0)$ Walker mode corresponds to that shown in Fig.\,\ref{Fig:Couplings}. The driving power is $15$ mW. Parameters in correspondence with the experiment \cite{potts2022dynamicalbackaction}, given in Table \ref{Table0}.}
\label{Fig:FreqShift}
\end{figure}

The magnomechanical frequency shift is also modified by the Kittel mode nonlinearity and by the coupling to the additional magnon mode. This is depicted in Fig.\,\ref{Fig:FreqShift}, which shows the magnomechanical frequency shift $\delta \Omega = - \rm{Re}[\Sigma_{\rm{Tot}} [\Omega_{\rm{b}}]]$ for (a) several values of the self-Kerr nonlinearity and (b) several values of the coupling to the additional magnon mode. Whereas the Kerr nonlinearity induced a tilt in the slope of the  magnomechanical decay rate, its effect on the magnomechanical frequency shift consists of an extra negative shift. We also notice that the magnomechanical frequency shift does not vanish for a drive at the frequency where the magnomechanical decay vanishes. This is the case because for the parameters considered, the Kittel mode frequency does not match the microwave frequency. For perfectly matching Kittel mode and microwave frequencies, and in the absence of additional magnon modes, both the magnomechanical decay and the magnomechanical frequency shift vanish at the same drive frequency \cite{potts2020magnon}.

\begin{figure}[H]
\includegraphics[width = \columnwidth]{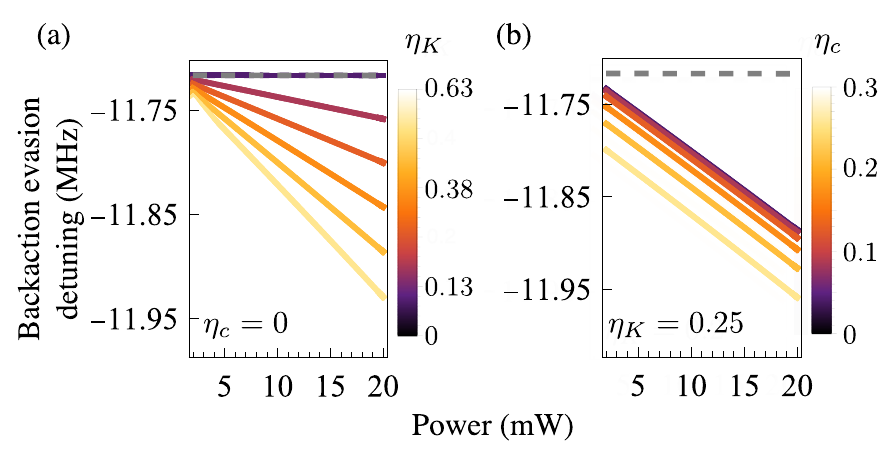}
\caption{Detuning between drive frequency and the upper hybrid mode for backaction evasion as a function of power for (a) no coupling to additional magnon modes and for several values of the dimensionless Kittel self-Kerr nonlinearity $\eta_{K}$, and (b) several values of the coupling between the $(4,3,0)$ Walker mode and the microwave cavity at a fixed $\eta_K = 0.25$. \hl{The dashed line corresponds to the prediction from the self-energy Eq .~\eqref{Eqn:06} derived in Ref.~\cite{potts2020magnon}}. Parameters in correspondence with the experiment \cite{potts2022dynamicalbackaction}, given in Table \ref{Table0}.}
\label{Fig:BAE}
\end{figure}

In Fig.~\ref{Fig:DecaysOne} one notices that the drive frequency at which the magnomechanical decay vanishes changes with both $\eta_K$ and $\eta_c$. For applications where evading backaction is important, it is necessary that such modifications are taken into account. We show in Fig.\,\ref{Fig:BAE} the drive frequency for backaction evasion (with respect to the upper hybrid mode frequency) as a function of the drive power for (a) several values of the Kittel self-Kerr nonlinearity and (b) several values of the coupling to the additional magnon mode at a fixed $\eta_K$. For the case without nonlinearities and without coupling to the additional magnon mode, the backaction evasion frequency has a weak dependency on power (not perceptible in the plot). When the corrections are included, a stronger linear dependency of the backaction drive frequency with the power is induced. For the parameters in consideration, the difference can be of the order of $\sim 0.1$ MHz at moderate powers of $10$ mW.

\begin{figure}[H]
\includegraphics[width = \columnwidth]{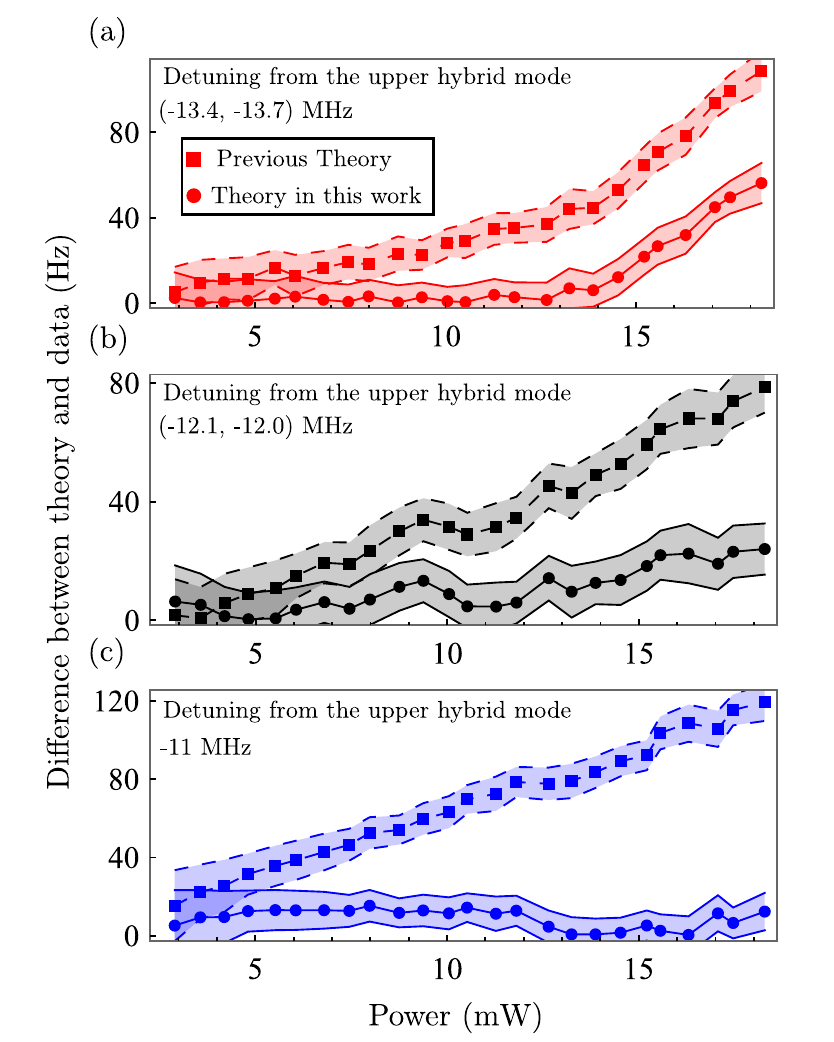}
\caption{Absolute difference between the theory for the magnomechanical decay and the experimental data as a function of power \hl{at the device} for a drive detuned from the upper hybrid mode by (a) $(-13.4, - 37.7)$ MHz, (b) $(-12.1, 12.0)$ MHz, and (c) $-11$ MHz. The dashed curves correspond to the prediction of the previous theory using \eqref{Eq:XiKit}, while the solid lines correspond to the theory developed in this paper. The shaded region corresponds to the experimental errors of the data obtained in Ref. \cite{potts2022dynamicalbackaction}. Theory predictions use parameters in correspondence with the experiment \cite{potts2022dynamicalbackaction}, given in Table \ref{Table0}.}
\label{Fig:Power}
\end{figure}

In order to quantify the agreement between our model and the measured data in Ref.~\cite{potts2022dynamicalbackaction}, we study the difference between the theoretical magnomechanical decay $\Gamma_{\rm{mag}}[\Omega_{\rm{b}}]$ rate and the experimental data $\Gamma_{\rm{exp}}$. In Fig.\,\ref{Fig:Power}, we show the absolute difference $\vert \Gamma_{\rm{mag}}[\Omega_{\rm{b}}] -\Gamma_{\rm{exp}} \vert$ between theory and experiment as a function of the drive power at the device for different drive frequencies. \hl{We should notice that in \cite{potts2022dynamicalbackaction}, there is a power loss of $\sim 2.38$ dBm between the generator and the device that has already been taken into account for this plot.} Our proposed model agrees well with the experimental data, besides the difference at higher powers and drives farther from the upper hybrid mode, as it is evident in Fig.~\ref{Fig:Power}(a). In the worst case, the model proposed here improves the discrepancy between data and theory from $\sim120$ Hz (dashed red curve in Fig.\,\ref{Fig:Power}), to a difference of $\sim50$ Hz (solid red curve in Fig.\,\ref{Fig:Power}). We also notice a significant overlap at low drive powers and for all the depicted detunings due to measurement errors. Otherwise, we notice good agreement between theory and experiment for drive powers up to $\sim 14$ mW. At such powers, the coherent number of magnons generated by the microwave drive $\vert \langle \hat{m} \rangle \vert^2$, see Eq.\,\eqref{App02:Eq05}, is between $\approx 6.0 \times 10^{13}$ at a detuning from the upper hybrid mode $ \Delta_{+} = -11$ MHz and $\approx  7.4 \times 10^{13}$ at $\Delta_{+} = -14$ MHz. We should notice that for the parameters considered here, the system is not in a bistable regime.

\begin{figure*}[t!]
  \includegraphics[width=  0.8 \textwidth]{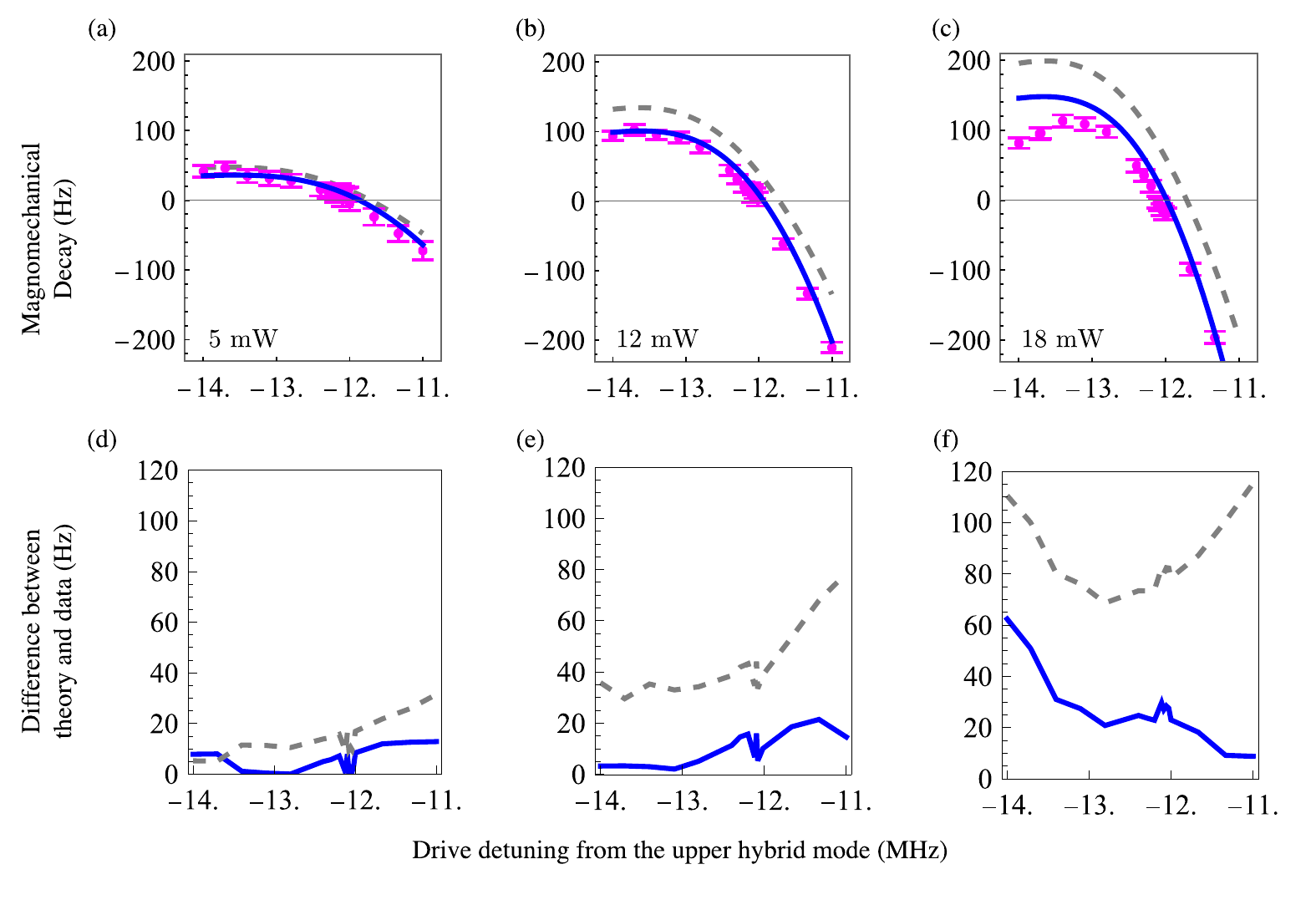}
  \caption{Comparison between the magnomechanical decay rate $\Gamma_{\rm{mag}}[\Omega_{\rm{b}}]$ predicted by Eq.~\eqref{Eqn:06} (dashed gray line), by Eq.~\eqref{App02:Eq17} (blue line) and the experimental data measured in Ref. \cite{potts2022dynamicalbackaction} (magenta points). In these plots we have used $\eta_c = 0.3$ and $\eta_K = 0.25$, which yields a good agreement between our model and the experimental data, especially close to the point of dynamical backaction evasion. (a)-(c): Magnomechanical decay rate as a function for the detuning; (d)-(f) absolute difference between the theory and the experiment as a function of the detuning (we have omitted the error bars in these plots for a better visualization). Theory curves with parameters in correspondence with the experiment \cite{potts2022dynamicalbackaction}, given in Table \ref{Table0}, \hl{and the driving powers indicated are at the device.}}
 \label{Fig:DecayComp}
\end{figure*}

 In Figs.~\ref{Fig:DecayComp} we show in (a-c) the magnomechanical decay as a function of the drive frequency detuning from the upper hybrid mode. \hl{The powers indicated in those figures are at the device and already take into account the $2.38$ dBm losses between the generator and the device.} As in the discussion above, we consider the coupling only to the $(4,3,0)$ Walker mode, and we choose $\eta_{\rm{K}}=0.25$ and $\eta_c = 0.3$, which yields a good agreement between theory and data. In the plots of Figs. \ref{Fig:DecayComp} (d-f), we show the difference $\vert \Gamma_{\rm{mag}}[\Omega_{\rm{b}}] -\Gamma_{\rm{exp}} \vert$. While the correction due to the coupling to the $(4,3,0)$ Walker mode improves the agreement between theory and data with respect to the previous theory framework \cite{potts2021dynamical}, we notice, as shown in Fig.~\ref{Fig:Power}, that at higher drive powers, there is a further discrepancy with the experiment for drives away from the dynamical backaction evasion points. We also notice that the errors of the data shown in Fig.~\ref{Fig:DecayComp} (a) do overlap with both the present theory and the one used in Ref. \cite{potts2021dynamical}, as it is also shown in Fig.~\ref{Fig:Power}.

\hl{We attribute the discrepancy at higher powers to other nonlinear phenomena that were not considered here. Specifically, the scattering of magnon modes into the spin wave continuum via three- and four- magnon processes are known to generate an instability of the spin waves, the Suhl instabilities \cite{clogston1956ferromagnetic, suhl1956subsidiaryabsorption, suhl1957thetheoryof}. Above a certain threshold drive power, the amplitude of the spin waves increases at the expense of a reduction in the amplitude of the magnon modes. This phenomenon scales with the drive power and is more prominent for drive frequencies close to the Kittel mode frequency. A power-dependent reduction in the average number of Kittel magnons would imply a power-dependent reduction of the magnomechanical decay rate, which, according to the requirements for the onset of the Suhl instabilities, would occur more prominently for drives at frequencies closer to the Kittel mode. Such effect is compatible with the behavior exhibited by the data, for example, in Fig.~\ref{Fig:DecayComp}, where a reduction in the magnomechanical decay happens for drive frequencies that are relatively close to the Kittel mode frequency (see the schematic in Fig.~\ref{Fig:MagnonMode}). In Ref. \cite{elyasi2020resources}, it was shown that the nonlinear behavior could occur for spheres of $100$ $\mu$m diameter at driving powers $\sim 1$ mW. Larger spheres, such as the ones considered here, would require a stronger driving power, but the drive frequency and powers at which the discrepancy is noticeable are compatible with the requirements for the onset of Suhl instabilities. A formal evaluation of such effects goes beyond the scope of this work and will be treated elsewhere.}

\section{Conclusions}
\label{Sec:Conclusions}

Dynamical backaction effects in magnomechanical systems are a consequence of the radiation pressure like coupling between magnons and phonons \cite{zhang2016cavity, potts2021dynamical}, which can be exploited for applications ranging from generating entangled states to noise-based thermometry \cite{potts2020magnon}. In this paper, we have extended the description of dynamical backaction in cavity magnomechanics by including in the system's dynamics self- and cross-Kerr nonlinearities, and the coupling between the phonon mode and additional magnon modes. While nonlinearities are intrinsic to magnetic systems due to, e.g., magnetic anisotropy \cite{stancil2009spinwaves}, magnon modes other than the uniform Kittel mode are always present and can couple to phonons as efficient as (if not more than) the Kittel mode. A nonuniform microwave field can weakly drive such modes, which modifies the backaction-induced decay rate and frequency shift of the phonon mode. Our framework considers a single phonon mode, an assumption that can be readily generalized.

We have obtained the phonon self-energy, including the aforementioned interactions and showed that, provided that the additional magnon modes couple only weakly to the microwave mode, the overall correction to the magnomechanical decay rate is proportional to the average number of Kittel magnons. We have then focused our results on the case of a magnetic sphere, in connection with the experiment performed in Ref.\,\cite{potts2022dynamicalbackaction}. Our model explains the observed shift in the magnomechanical decay rate close to the dynamical backaction evasion drive frequency. In this context, we have also evaluated the effects of the different corrections. Specifically, we showed that the drive at which the dynamical backaction decay is zero depends linearly on power. This is a consequence of the corrections being proportional to the steady-state number of Kittel magnons, which scales linearly with the drive power.

A small discrepancy with the experimental data is still present at higher drive powers and for detunings far from the upper hybrid mode. We attribute this difference to \hl{nonlinear processes, such as scattering into spin waves generating Suhl instabilities \cite{clogston1956ferromagnetic,suhl1956subsidiaryabsorption, suhl1957thetheoryof, elyasi2020resources}.} Furthermore, even higher-order Walker modes can lie in a frequency range close to the microwave drive and, at higher powers, can modify the magnomechanical decay. \hl{Preliminary calculations including five more Walker modes have shown an incremental improvement of our model.} We should also point out that the experimental setup in Ref. \cite{potts2022dynamicalbackaction} has particularities not included here. For instance, the magnetic sphere is glued on a dielectric post, which modifies the photon, phonon, and magnon mode profiles. This in turn can change the magnomechanical coupling constants as well as the frequency of the Walker modes. A precise evaluation of such effects requires a more refined numerical analysis, for example, using finite difference software and micromagnetic simulations, which goes beyond the scope of our analysis. 

While nonlinear effects in cavity magnomechanical systems have been previously computed for the nonlinear dynamics of magnons \cite{shen2022mechanical,elyasi2020resources}, the evaluation of such effects on the response of the mechanical degree of freedom to noise, as computed by the self-energy, is a step forward in the characterization of these systems as platforms for quantum technologies. Our analysis was restricted to evaluate the effects of all the corrections included in the model of Sec.\,\ref{Sec:AdditionalModesSelfEnergy} in the framework of dynamical backaction evasion set by the experiment of Ref.\,\cite{potts2022dynamicalbackaction}. Nevertheless, the model derived in Sec.\,\ref{Sec:AdditionalModesSelfEnergy} shows that several phenomena play a role in the modification of dynamical backaction, for example, magnon squeezing and two-mode squeezing. It would be interesting to investigate scenarios in which those terms can be harnessed to reduce noise for quantum metrology. Furthermore, the inclusion of the additional magnon modes opens new possibilities for cavity magnomechanical systems, such as the manipulation of the mechanics by driving different sidebands of the different magnon modes in a Floquet-like setup \cite{xu2020floquetcavity}. In this case, it would be interesting to go beyond the approximation used here, where only the Kittel mode couples strongly to the microwave cavity. In fact, several experiments have shown fingerprints of a strong coupling between Walker modes of a sphere and microwaves \cite{zhang2015cavityquantum, wang2016magnonkerr, morris2017strong}. As we have numerically shown, Walker modes other than the Kittel mode can couple better to the phonons, which can be harnessed to applications, such as nonreciprocal transport between phonons and microwaves \cite{lepinay2020nonreciprocal}.

\begin{acknowledgments}

The authors acknowledge helpful contributions from S. Scharma and E. Varga.  V.A.S.V. Bittencourt and S. Viola Kusminskiy acknowledge financial support from the Max Planck Society and from the Deutsche Forschungsgemeinschaft (DFG, German Research Foundation) through Project-ID 429529648–TRR 306 QuCoLiMa (“Quantum Cooperativity of Light and Matter”). C.A. Potts, Y. Huang, and J.P. Davis acknowledge support by the University of Alberta; the Natural Sciences and Engineering Research Council, Canada (Grant Nos. RGPIN-2022-03078, and CREATE-495446-17); the NSERC Alberta Innovates Advance program; and the Government of Canada through the NRC Quantum Sensors Program.
\end{acknowledgments}

\appendix
\hl{
\section{Formulas omitted in the main text}
\label{AppA}
In the following we present the formulas that were omitted in the main text.

The modified coupling rates appearing in Eq.~\eqref{App02:Eq12} are given by
\begin{widetext}
\begin{equation}
\label{Eq:ModfCoup}
\begin{aligned}
\tilde{G}_{\rm{b,R}}[\omega] &= G_{\rm{R}} - g_{\rm{am}} \chi_{a}^*[-\omega] \sum_{j} g_{\rm{a m_j}} G_{\rm{R,j}} \Xi_j^*[-\omega] - i \sum_{j} \left(G_{\rm{B,j}} g_{\rm{b,j}} \Xi_j[\omega]-G_{\rm{R,j}} g_{\rm{R,j}}  \Xi_j^*[-\omega]   \right), \\
\tilde{G}_{\rm{b,B}}[\omega] &= G_{\rm{B}} -  g_{\rm{am}} \chi_{\rm{a}}^*[-\omega] \sum_j g_{\rm{a m_j}} G_{\rm{B,j}}^* \Xi_j^*[-\omega] - i \sum_{j} \left( G_{\rm{R,j}}^* g_{\rm{b,j}} \Xi_j[\omega] - G^*_{\rm{B,j}} g_{\rm{R,j}}  \Xi_j^*[-\omega] \right),\\
G_{\rm{m,R}}[\omega] &= G_{\rm{R}} - g_{\rm{am}} \chi_{\rm{a}}[\omega] \sum_{j} g_{\rm{a m_j}} G_{\rm{R,j}} \Xi_j [\omega] - i \sum_j \left(g_{\rm{R,j}} G_{\rm{R,j}} \Xi_j [\omega] - g_{\rm{B,j}} G_{\rm{B,j}} \Xi_j^* [-\omega] \right),\\
G_{\rm{m,B}}[\omega] &= G_{\rm{B}} - g_{\rm{am}} \chi_{\rm{a}}[\omega] \sum_{j} g_{\rm{a m_j}} G_{\rm{B,j}}^* \Xi_j [\omega] - i \sum_j \left(g_{\rm{R,j}} G_{\rm{B,j}}^* \Xi_j [\omega] - g_{\rm{B,j}} G_{\rm{R,j}}^* \Xi_j^* [-\omega] \right).
\end{aligned}
\end{equation}
\end{widetext}
The second terms in Eqs.~\eqref{Eq:ModfCoup} represent the effect of the indirect coupling between the additional magnon modes and the Kittel mode via the cavity. The third terms are proportional to $g_{\rm{B(R),j}}$, which in turn (see Table \ref{Table01}) are due to the magnon cross-Kerr nonlinearity, $\hat{m}_j^\dagger \hat{m}_j \hat{m}^\dagger \hat{m}$ in the Hamiltonian of Eq.~\eqref{App02:Eq01}. 

The modified magnomechanical couplings appearing in the final formula for the phonon self-energy in Eq.~\eqref{App02:Eq16} are given by
\begin{equation}
\begin{aligned}
\tilde{G}_{\rm{m,R}}[\omega] &= G_{\rm{m,R}}[\omega] + i \frac{\Lambda_{\rm{m}} [\omega]  G_{\rm{m,B}}^*[-\omega]}{\Xi_{\rm{m}} ^{*,-1}[-\omega]-\eta_{\rm{m}} ^*[-\omega]},  \\
\tilde{G}_{\rm{m,B}}[\omega] &=  G_{\rm{m,B}}[\omega] + i \frac{\Lambda_{\rm{m}} [\omega]  G_{\rm{m,R}}^*[-\omega]}{\Xi_{\rm{m}} ^{*,-1}[-\omega]-\eta_{\rm{m}} ^*[-\omega]}. 
\end{aligned}
\end{equation}
We see that the effective coupling rates $G_{\rm{m,R (B)}}[\omega]$, c.f. Eq.~\eqref{Eq:ModfCoup}, which include only contributions due to the indirect coupling between the Kittel mode and the additional magnon modes, are further modified by the Kittel mode squeezing term $\Lambda_{\rm{m}}[\omega]$. 

\section{Magnomechanical coupling}
\label{AppB}

We first quantize the magnetization. We consider that the magnetization displays small fluctuations around a uniform saturation value $M_S$ \cite{stancil2009spinwaves} (such assumption can be generalized to nonuniform magnetic ground states \cite{graf2018cavityoptomagnonics}). In this framework $M_z/M_S \gg M_{x,y}/M_S$. The $x$ and $y$ components of the magnetization can then be written as a superposition of modes, each labeled with a general index $j$, such that the quantized magnetization field is given by \cite{Mills_2006}
\begin{equation}
\label{Eq:QuantMagnetization}
\hat{M}_{x,y}(\bm{r},t) = \sum_{j}\mathcal{M}_{j} \left[ \delta m_{x,y;j} (\bm{r}) \hat{m}_j + \delta m_{x,y;j}^* (\bm{r}) \hat{m}^\dagger_j \right],
\end{equation}
where $\{\hat{m}_j\}$ is a set of bosonic operators satisfying $[\hat{m}_j, \hat{m}^\dagger_j]=1$. The quantization procedure is valid in the so-called spin-wave limit for magnetic excitations. The mode functions $\delta \bold{m}_{j} (\bm{r})$ are obtained by solving the Landau-Lifshitz equation for the magnetization fluctuations \cite{stancil2009spinwaves}, plus the appropriate boundary conditions. The quantities $\mathcal{M}_{j}$ are the zero-point fluctuations of the mode $j$ given by
\begin{equation}
\mathcal{M}_{j} = \sqrt{\frac{\hbar \vert \gamma \vert M_{\rm{S}}}{2 V_{j}}},
\end{equation}
where the mode volume is given by \cite{Mills_2006}
\begin{equation}
V_{j} = 2 {\rm{Im}}\left[\int d^3 r \, \delta m_{y;j} (\bm{r}) \delta m_{x;j}^* (\bm{r})  \right].
\end{equation}
Such mode decomposition ensures that the magnetic energy density yields the Hamiltonian for a set of uncoupled harmonic oscillators of the form $\mathcal{\hat{H}}_{\rm{m}} = \sum_j \hbar \omega_j \hat{m}_j^\dagger \hat{m}_j$, with frequencies $\omega_j$ obtained from the imposed boundary conditions.

The elastic vibrations are quantized in terms of phonon modes \cite{anghel2007quantization}. The displacement field is given by the superposition of modes
\begin{equation}
\label{Eq:QuantDisplacement}
\hat{\bm{u}} = \sum_{\alpha} \mathcal{X}_{\alpha} \left[ \bm{f}_{\alpha}(\bm{r}) \hat{b}_\alpha + \bm{f}^*_{\alpha}(\bm{r}) \hat{b}^\dagger_\alpha\right].
\end{equation}
The mode functions $\bm{f}_{\alpha}(\bm{r})$ are dimensionless, and given as the solution of the elastic boundary problem \cite{eringen1975elastodynamics}.
The zero-point fluctuations are given by
\begin{equation}
\mathcal{X}_{\alpha} = \sqrt{\frac{\hbar }{2 \rho \Omega_\alpha N_\alpha}},
\end{equation}
where
\begin{equation}
  N_\alpha = \int d^3 r \, \vert  \bm{f}_{\alpha}(\bm{r}) \vert^2
\end{equation}
is the mode normalization. Such a mode decomposition yields, for the noninteracting phonons, the Hamiltonian $\mathcal{\hat{H}}_{\rm{b}} = \sum_\alpha \Omega_\alpha \hat{b}^\dagger_\alpha \hat{b}_\alpha$.

Substituting Eqs.~\eqref{Eq:QuantMagnetization} and \eqref{Eq:QuantDisplacement} in the magnetoelastic energy given by Eq.\,\eqref{Eq:MagnetoElastic}, we obtain an interaction Hamiltonian describing the coupling between magnons and phonons. Such a Hamiltonian includes the following terms: (i) linear magnon-phonon coupling $\propto g_{m_j b_\alpha}^{(L)} \hat{m}_j^\dagger \hat{b}_\alpha + \rm{H.c.}$, relevant only for resonant magnon and phonon modes, for example, for small magnetic particles \cite{Gonzalez_2020_Theory} and for magnetic films 
\cite{an2020coherentlongrange, litvinenko2021tunablemagnetoacoustic,schlitz2022magnetizationdynamics}; (ii) spontaneous parametric conversion terms $\propto  \hat{m}_j \hat{m}_k \hat{b}_\alpha^\dagger$ and $\propto  \hat{m}_j^\dagger \hat{m}_k^\dagger \hat{b}_\alpha$, relevant when the phonon mode frequency matches the sum of the frequency of the magnon modes $j$ and $k$, where such an interaction describes the creation of a pair of magnons via the annihilation of a phonon; and (iii) parametric phonon-magnon coupling $\propto g_{\rm{m_j m_k b_\alpha}}^0 \hat{m}_j^\dagger \hat{m}_k \hat{b}+ \rm{H.c}$. The off-resonant terms $\hat{m}_j \hat{m}_k \hat{b}_\alpha$ and $\hat{m}_j^\dagger \hat{m}_k^\dagger \hat{b}_\alpha^\dagger$ can be eliminated via a rotating wave approximation. 

Our focus is on the parametric interaction (iii), for which the Hamiltonian is given by Eq.~\eqref{Eq:GenParHamilt}:
\begin{equation}
\begin{aligned}
\hat{\mathcal{H}}_{\rm{mb}}/\hbar &=\sum_{\{j\neq k\}, \alpha} \left[g_{\rm{m_k  m_j  b_\alpha}}^{0} \hat{m}_k^\dagger \hat{m}_j \hat{b}_\alpha + \tilde{g}_{\rm{m_k  m_j  b_\alpha}}^{0} \hat{m}_k^\dagger \hat{m}_j \hat{b}^\dagger_\alpha  \right]  \\
&\quad + \sum_{j, \alpha} g_{\rm{m_j  b_\alpha}}^{0} \hat{m}_j ^\dagger \hat{m}_j \hat{b}_\alpha + \rm{H.c.},
\end{aligned}
\end{equation}
where the coupling rates are given by
\begin{widetext}
\begin{equation}
\label{Eq:Couplings}
\begin{aligned}
\frac{g_{\rm{ m_j  b_\alpha}}^0}{\mathcal{N}_{\rm{ m_j  b_\alpha}}} &= B_1 \int d^3 r \, \left[ \vert \delta m_{x;j}(\bm{r})\vert^2 (\partial_x f_{x;\alpha}(\bm{r}) - \partial_z f_{z;\alpha}(\bm{r})) + \vert \delta m_{y;j}(\bm{r})\vert^2 (\partial_x f_{x;\alpha}(\bm{r}) - \partial_z f_{z;\alpha}(\bm{r}))\right] \\
&\quad + B_2 \int d^3 r \, {\rm{Re}}\left[ \delta m_{x;j}(\bm{r}) \delta m_{y;j}^*(\bm{r}) \right] \left( \partial_y f_{x;\alpha}(\bm{r})+\partial_x f_{y;\alpha}(\bm{r}) \right), \\
\frac{g_{\rm{m_k  m_j  b_\alpha}}^0}{\mathcal{N}_{\rm{m_k  m_j  b_\alpha}}} &= B_1 \int d^3 r \, \left[ \delta m_{x;k}^*(\bm{r}) \delta m_{x;j}(\bm{r}) (\partial_x f_{x;\alpha}(\bm{r}) - \partial_z f_{z;\alpha}(\bm{r})) + \delta m_{y;k}^*(\bm{r}) \delta m_{y;j}(\bm{r}) (\partial_y f_{y;\alpha}(\bm{r}) - \partial_z f_{z;\alpha}(\bm{r})) \right] \\
&\quad + \frac{ B_2 }{2} \int d^3 r \,\left[ \delta m_{y;k}^*(\bm{r}) \delta m_{x;j}(\bm{r}) +\delta m_{x;k}^*(\bm{r}) \delta m_{y;j}(\bm{r}) \right] \left( \partial_y f_{x;\alpha}(\bm{r})+\partial_x f_{y;\alpha}(\bm{r}) \right),
\end{aligned}
\end{equation}
\end{widetext}
where we have defined
\begin{equation}
\begin{aligned}
\mathcal{N}_{\rm{m_k  m_j  b_\alpha}}=2 \frac{\mathcal{X}_\alpha \mathcal{M}_k \mathcal{M}_j}{\hbar M_{\rm{S}}^2},
\end{aligned}
\end{equation}
and $\mathcal{N}_{\rm{ m_j  b_\alpha}}=\mathcal{N}_{\rm{m_j  m_j  b_\alpha}}$. The coupling $\tilde{g}_{\rm{m_k  m_j  b_\alpha}}^{0}$ is obtained from $g_{\rm{m_k  m_j  b_\alpha}}^{0}$ with the substitution $\partial_{x_i}f_{j;\alpha}(\bm{r}) \rightarrow \partial_{x_i}f^*_{j;\alpha}(\bm{r})$.

\section{Magnetostatic and mechanical modes of a sphere}
\label{AppC}

A sphere supports magnetostatic modes called Walker modes \cite{walker1958resonant,fletcher1959ferrimagnetic, roschmann1977propertiesof}, which have frequencies that can be tuned by the value of the external bias field. To describe such modes, it is convenient to introduce the following characteristic frequencies: 
\begin{equation}
\begin{aligned}
\omega_{\rm{M}} &= \vert \gamma \vert \mu_0 M_{\rm{S}}, \\
\omega_0 &= \vert \gamma \vert \mu_0 \left(H_0 - \frac{M_{\rm{S}}}{3} \right),
\end{aligned}
\end{equation}
where $\vert \gamma \vert/ 2 \pi = 28$ GHz/T is the gyromagnetic ratio, $\mu_0$ is vacuum permeability, and $H_0$ is the applied bias magnetic field. The Walker modes are conveniently given in a nonorthogonal coordinate system $\{\xi, \eta, \phi \}$ defined by the transformation \cite{walker1958resonant}
\begin{equation}
\label{Eq:WalkerCoords}
\begin{aligned}
x &= R \sqrt{-\chi_{\rm{P}}[\omega]} \sqrt{1 - \xi^2} \sin{\eta} \cos{\phi}, \\
y &= R \sqrt{-\chi_{\rm{P}}[\omega]} \sqrt{1 - \xi^2} \sin{\eta} \sin{\phi},  \\
z &= R \sqrt{\frac{\chi_{\rm{P}}[\omega]}{1+\chi_{\rm{P}}}} \xi \cos{\eta},
\end{aligned}
\end{equation}
where 
\begin{equation}
\chi_{\rm{P}}[\omega]= \frac{\omega_{M} \omega_0}{\omega_0^2 - \omega^2}.
\end{equation}
At the the sphere's surface $\eta \rightarrow \theta$ and 
\begin{equation}
\xi[\omega] \rightarrow \xi_0[\omega] = \sqrt{\frac{1+\chi_{\rm{P}}[\omega]}{\chi_{\rm{P}}[\omega]}}.
\end{equation}
The frequencies of Walker modes are given by the nonlinear equation \cite{walker1958resonant,fletcher1959ferrimagnetic}
\begin{equation}
\label{WalkerFrequencies}
\xi_0[\omega] \frac{\partial_{\xi} P^m_l (\xi[\omega])}{P^m_l (\xi[\omega])}\vert_{\xi = \xi_0} - m \kappa_{\rm{P}}[\omega] + n + 1 =0,
\end{equation}
where
\begin{equation}
\kappa_{\rm{P}}[\omega] = -\frac{\omega_{M} \omega}{\omega_0^2 - \omega^2},
\end{equation}
and $P^m_l$ are the associated Legendre polynomials. The Walker modes are labeled by three indices, $\{ l m \nu \}$, with $l\ge 1$ and $\vert m \vert \le l$. For $m>0$, Eq. \eqref{WalkerFrequencies} has $(l - \vert m \vert)/2$ roots, while for $m<0$ it has $1+(l - \vert m \vert)/2$ solutions (both rounded down). The mode functions of the Walker modes are given by
\begin{equation}
\begin{bmatrix}
\delta m_{x;l m \nu}\\
\delta m_{y;l m \nu}
\end{bmatrix}
=
- \begin{bmatrix}
 \chi_{\rm{P}} [\omega_{l m \nu}]  & i \kappa_{\rm{P}}[\omega_{l m \nu}] \\
-i \kappa_{\rm{P}}[\omega_{l m \nu}] &  \chi_{\rm{P}} [\omega_{l m \nu}]
\end{bmatrix} \begin{bmatrix}
\partial_x \psi_{l m \nu }\\
\partial_y \psi_{l m \nu }
\end{bmatrix}
\end{equation}
where the magnetostatic potential inside the sphere is
\begin{equation}
\psi_{l m \nu }(\bm{r}) = P^{m}_l (\xi) Y^m_l (\eta, \phi).
\end{equation}

For the phonon modes, we consider an unpinned sphere and stress-free boundary conditions \cite{eringen1975elastodynamics}. There are two families of mechanical modes of a homogeneous sphere: torsional (T) and spherical (S) modes. Torsional modes are purely shear modes, while spherical modes involve both shear and compression. Both families of modes are labeled by three indices $\{\nu l m\}$, where $l$ and $m$ are polar and azimuthal indices $-l \le m \le l$ while $\nu$ is a radial index. We focus here on S modes, whose frequencies are given by \cite{eringen1975elastodynamics}
\begin{equation}
\label{Eq:FreqPhon}
\mathcal{T}^{(a)}_{\lambda \nu} \mathcal{T}^{(b)}_{\lambda \nu} - \mathcal{T}^{(c)}_{\lambda \nu} \mathcal{T}^{(d)}_{\lambda \nu} =0,
\end{equation}
where
\begin{equation}
\begin{aligned}
\mathcal{T}^{(a)}_{\lambda \nu} &= \left[\lambda (\lambda -1) - \frac{\tilde{\beta}^2[\omega] R^2}{2} \right]j_{\lambda}(\tilde{\alpha}[\omega] R) \\
&\quad+ 2 \tilde{\alpha}[\omega] R j_{\lambda+1}(\tilde{\alpha}[\omega] R), \\
\mathcal{T}^{(b)}_{\lambda \nu} &= \left[\lambda^2 -1 - \frac{\tilde{\beta}^2[\omega] R^2}{2} \right]j_{\lambda}(\tilde{\beta}[\omega] R) \\
&\quad +  \tilde{\beta}[\omega] R j_{\lambda+1}(\tilde{\beta}[\omega] R), \\
\mathcal{T}^{(c)}_{\lambda \nu} &= \lambda (\lambda + 1)\Big[ (\lambda - 1) j_{\lambda}(\tilde{\beta}[\omega] R), \\
&\quad \quad \quad \quad \quad \quad - \tilde{\beta}[\omega] R j_{\lambda+1}(\tilde{\beta}[\omega] R) \Big] \\
\mathcal{T}^{(d)}_{\lambda \nu} &= (\lambda - 1) j_{\lambda}(\tilde{\alpha}[\omega] R) - \tilde{\alpha}[\omega] R j_{\lambda +1}(\tilde{\alpha}[\omega] R).
\end{aligned}
\end{equation}
The parameters $\tilde{\alpha}[\omega]= \omega/c_L$, and $\tilde{\beta}[\omega]= \omega/c_T$, are given in terms of the longitudinal (L) and transverse (T) sound velocities $c_{L,T}$. $j_{\lambda}(x)$ denotes the spherical Bessel function. Since Eq. \eqref{Eq:FreqPhon} does not depend on $m$, for given $\{\nu l\}$ there are $2l +1$ degenerate modes. The mode functions for an S mode read, in spherical coordinates $\{\bm{e}_r, \bm{e}_{\theta}, \bm{e}_\phi \}$,
\begin{equation}
\bm{f}_{\nu \lambda m}= e^{i \phi m} \begin{bmatrix}
 \mathcal{G}_{\nu \lambda} (r) P^m_l (\cos \theta) \\
 \mathcal{F}_{\nu \lambda} (r) \partial_\theta P^m_l (\cos \theta) \\
 \frac{i m}{ \sin \theta}\mathcal{F}_{\nu \lambda} (r) P^m_l (\cos \theta)
\end{bmatrix},
\end{equation}
where
\begin{equation}
\begin{aligned}
 \mathcal{G}_{\nu \lambda} (r)&= \frac{R}{r} \Big[ \lambda j_{\lambda}(\tilde{\alpha}[\omega] r) - \tilde{\alpha}[\omega] r j_{\lambda +1}(\tilde{\alpha}[\omega] r)  \\
 &\quad \quad \quad \quad \quad - \frac{\mathcal{T}^{(d)}_{\lambda \nu}}{\mathcal{T}^{(b)}_{\lambda \nu}} \lambda (\lambda+1)j_{\lambda}(\tilde{\beta}[\omega] r)  \Big],   \\
 \mathcal{F}_{\nu \lambda} (r)&= \frac{R}{r} \Big[j_{\lambda}(\tilde{\alpha}[\omega] r) + \frac{\mathcal{T}^{(d)}_{\lambda \nu}}{\mathcal{T}^{(b)}_{\lambda \nu}} \tilde{\beta}[\omega] r j_{\lambda+1}(\tilde{\beta}[\omega] r) \\
 &\quad \quad \quad \quad \quad -\frac{\mathcal{T}^{(d)}_{\lambda \nu}}{\mathcal{T}^{(b)}_{\lambda \nu}} (\lambda+1)j_{\lambda}(\tilde{\beta}[\omega] r)\Big].
 \end{aligned}
\end{equation}
The experiment in Ref. \cite{potts2022dynamicalbackaction} has probed the coupling to the $S_{122}$ mode, which is our case of study throughout the main text.

\section{Phases of the magnomechanical couplings}
\label{AppD}

The magnomechanical coupling rates appearing in Eq.~\eqref{Eq:MagnomechHi} are complex numbers, but we can absorb the phase of one of such coupling rate into the phonon field. Specifically, we chose to absorb the phase of the Kittel mode magnomechanical coupling. We write $g_{\rm{mb}}^0 = \vert g_{\rm{mb}}^0 \vert e^{i \phi_{\rm{mb}}}$, and define $\hat{\tilde{b}} = \hat{b} e^{i \phi_{\rm{mb}}}$, such that
\begin{equation}
\label{Hamiltonian0}
\begin{aligned}
\frac{\hat{\mathcal{H}}_{\rm{m\tilde{b}}}}{\hbar} &= \omega_{\rm{m}} \hat{m}^\dagger \hat{m}+ \sum_{j} \omega_{j} \hat{m}_{j}^\dagger \hat{m} + \Omega_{\rm{b}} \hat{\tilde{b}}^{\dagger} \hat{\tilde{b}} \\
&\quad + g_{ \rm{m \tilde{b}}}^0 \hat{m}^\dagger \hat{m} (\hat{\tilde{b}} +\hat{\tilde{b}}^{\dagger}) \\
&\quad + \sum_{j} \left[ g_{\rm{m_j \tilde{b}}}^0 \hat{m}_j^\dagger \hat{m}_j \hat{\tilde{b}} + {\rm{H.c.}} \right] \\
&\quad + \hat{m}^\dagger \sum_j \left[ g_{\rm{m m_j \tilde{b} }}^0 \hat{m}_{j} \hat{\tilde{b}} + {\rm{H.c.}} \right] \\
&\quad +  \sum_{j\neq k} \left[ g_{\rm{m_k m_j \tilde{b} }}^0 \hat{m}_k^\dagger \hat{m}_{j} \hat{\tilde{b}} + {\rm{H.c.}} \right],
\end{aligned}
\end{equation}
where $g_{ \rm{m\tilde{b}}}^0=\vert  g_{ \rm{mb}}^0 \vert $, $ g_{\rm{m_j \tilde{b}}}^0 = g_{\rm{m_j b}}^0 e^{-i \phi_{\rm{mb}}}$, and $g_{\rm{m m_j \tilde{b} }}^0 = g_{\rm{m m_j b }}^0 e^{-i \phi_{\rm{mb}}}$. Such a transformation corresponds to taking the phase of the coupling between the phonon mode and the Kittel mode as a reference for the other couplings. From now on, we take $\hat{\tilde{b}} \rightarrow \hat{b}$. This gauge transformation of the phonon field does not change the Kerr nonlinear terms, which are quadratic in the phonon field. 

}

\bibliography{apssamp}

\end{document}